\documentclass[twocolumn]{aastex631}
\usepackage{textcomp}
\usepackage{url}
\usepackage{placeins}
\usepackage{multirow}
\usepackage{grffile}
\usepackage{amsmath}
\usepackage{amssymb,latexsym}
\usepackage{threeparttable}
\usepackage{longtable}
\usepackage[caption=false]{subfig}
\usepackage{ulem} 
\usepackage{natbib}
\usepackage{apjfonts}
\usepackage{mathtools}
\usepackage[inline]{asymptote}
\usepackage{xltabular}
\usepackage{booktabs}
\usepackage{hyperref}
\usepackage{verbatim}
\usepackage{threeparttable}

\hypersetup{
     colorlinks   = true,
     citecolor     = blue
}

\providecommand{\feh}{\ensuremath{\left[{\rm Fe}/{\rm H}\right]}}
\providecommand{\teff}{\ensuremath{T_{\rm eff}}}

\providecommand{\msun}{\ensuremath{\,M_\Sun}}
\providecommand{\rsun}{\ensuremath{\,R_\Sun}}

\newcommand\msini{\ifmmode{{\mathrm M} \sin i}\else${{\mathrm M} \sin i}$\fi}

\shorttitle{Direct Imaging Yield}
\shortauthors{Plavchan et al.}

\begin{document}

\title{Analytic relations assessing the impact of precursor knowledge and key mission parameters on direct imaging survey yield}

\author[0000-0002-8864-1667]{Peter Plavchan}
\correspondingauthor{Peter Plavchan, pplavcha@gmu.edu}
\affil{Department of Physics \& Astronomy, George Mason University, 4400 University Drive MS 3F3, Fairfax, VA 22030, USA}

\author[0000-0003-1466-8389]{John E. Berberian Jr.}
\affil{Department of Physics \& Astronomy, George Mason University, 4400 University Drive MS 3F3, Fairfax, VA 22030, USA}
\affil{Carter G. Woodson High School, 9525 Main St, Fairfax, VA 22031, USA}
\affil{University of Virginia, Charlottesville, VA, USA}

\author[0000-0002-7084-0529]{Stephen R. Kane}
\affil{Department of Earth and Planetary Sciences, University of California, Riverside, CA 92521, USA}

\author[0000-0002-4852-6330]{Rhonda Morgan}
\affil{Jet Propulsion Laboratory, California Institute of Technology, 4800 Oak Grove Dr, Pasadena, CA 91011, USA}

\author[0000-0003-2032-3336]{Eliad Peretz}
\affil{NASA Goddard Space Flight Center, Greenbelt, MD, USA}

\author[0000-0002-5534-3591]{Sophia Economon}
\affil{Florida Institute of Technology, 150 W University Blvd, Melbourne, FL 32901} 
\affil{Morton K. Blaustein Department of Earth and Planetary Sciences, John Hopkins University, 301 Olin Hall, 3400 N. Charles Street, Baltimore, MD 21218}



\begin{abstract}

The Habitable Worlds Observatory will attempt to image Earth-sized planets in Habitable Zone orbits around nearby Sun-like stars. In this work we explore approximate analytic yield calculations for a future flagship direct imaging mission for a survey sample of uniformly distributed set of identical Sun-like stars. We consider the dependence of this exoplanet detection yield on factors such as $\eta_\oplus$, telescope diameter, total on-sky time, orbital phase and separation, inner working angle, flux contrast, desired signal-to-noise ratio, spectral resolution, and other factors. We consider the impact on yield and survey efficiency in the absence of and with precursor knowledge of the Earth-size analog exoplanets.  In particular, for precursor knowledge we assume the exoplanet orbital phase at the time of observation can be optimized so as to only image the Earth-size analog exoplanet when it is outside the inner working angle.  We find that the yield of flagship direct imaging missions such as Habitable Worlds Observatory will be inner-working angle limited for the estimated exoplanet yields, and will not be impacted by precursor knowledge given our assumptions presented herein.  However, we find that the survey efficiency will be enhanced by precursor knowledge. We benchmark our analytic approximations against detailed simulations for coronagraphs and starshades carried out for the HabEx and LUVOIR missions concept studies, and find consistent conclusions.  Our analytic relations thus provide quick estimates and derivatives of the impact of key mission parameter choices on exo-Earth yield when considering design trades that can supplement existing computational simulations.

\end{abstract}



\section{Introduction}
\label{intro}

Over the past three decades, more than 5500 exoplanets have been discovered to orbit other stars, and the pace of discovery is accelerating \citep{Akeson_2013}.  As time has progressed, the main methods for exoplanet detection have been continually refined and improved to increase sensitivity to smaller and less massive planets orbiting main sequence stars.  The method of exoplanet direct imaging was first successful in imaging 2MASS 1207 b in \citet{2004A&A...425L..29C}, with now over 60 exoplanets discovered via direct imaging\citep{Akeson_2013}, including the multiplanet system HR 8799 \citep{2008Sci...322.1348M}. Prior to the 2020 Astrophysics Decadal Survey, NASA undertook the study of four flagship mission concept studies led by four science and technology definition teams (STDTs), which produced reports submitted for consideration by the Decadal Survey  \citep{gaudi2020habitable, theluvoirteam2019luvoir,OST2019,2019JATIS...5b1001G}.  Two of these mission concepts, HabEx and LUVOIR, considered the possibility of imaging other Earth-sized planets orbiting in the Habitable Zones (HZ) around nearby, Sun-like stars (hereafter exo-Earths) \citep{1993Icar..101..108K,2013ApJ...765..131K,2014ApJ...787L..29K,2016ApJ...830....1K}.  The 2020 Astrophysics Decadal Survey has recommended the development program for a future flagship direct imaging mission with a primary mirror of $\sim$6 m \citep{NAP26141}.  NASA has in turn launched the GOMAP (Great Observatory MAturation Program) and the START (Science, Technology, Architecture Review Team) for the HWO (Habitable Worlds Observatory)\footnote{https://science.nasa.gov/astrophysics/programs/gomap/}. 

Central to the scientific motivation for the Habitable Worlds Observatory, and predecessor mission concepts HabEx and LUVOIR, is the yield or number of exo-Earths these missions could be able to detect and characterize. As part of evaluating the feasibility of these mission concepts, detailed numerical simulations have been carried out to assess the yield of directly imaged exoplanets. In particular, a Standard  Definitions and Evaluation Team was formed by NASA with joint members from the mission concepts to evaluate mission yields, including such definitions as a common standard for the assumed exoplanet demographics, the location of the Habitable Zone and exoplanet size categories \citep{Standards,Dulz,kopparapu}. 
One of the key quantities that drives these mission yields is the exo-Earth occurrence rate  $\eta_\oplus$; the smaller this value, the larger a telescope will be needed. Direct imaging mission studies place much focus on understanding the impact of $\eta_\oplus$ and its corresponding uncertainty on the yield. Recent estimates of $\eta_\oplus$ from the Kepler mission have declined but also  increased in precision and knowledge \citep[e.g., ][and references therein]{2021AJ....161...36B,2019MNRAS.483.4479Z}, due to improvements of our understanding of the reliability and completeness of the Kepler mission exoplanet search, and also due to improvements in knowledge of our stellar parameters, particularly the stellar radius \citep{2014PASP..126...34P} with the release of Gaia DR2 and eDR3 \citep{2016A&A...595A...1G,2018AJ....156...58B,2018A&A...616A...1G,2021A&A...649A...1G}.  Many other factors impact mission yield and design, including astrophysical considerations such as the actual distribution of nearby stars and spectral types and exoplanet system demographics, and also mission parameters such as requirements on spectral resolution and grasp, signal to noise ratio, overhead time, assumed flux contrast ratio, etc. as explored in detailed simulations in \citet{Stark_2014,Stark_2015} and \citet{2021SPIE.0021215}.  The HabEx and LUVOIR mission concepts also considered the possibility that precursor knowledge of the existence, or lack thereof, could impact mission yield, such as could be provided by radial velocities or astrometry.  In particular for LUVOIR B and the three HabEx mission concepts, including those with a starshade, \citet{morgan} carried out detailed simulations assessing this impact.  They found that while precursor knowledge had a minor impact on mission yield, it did significantly impact survey efficiency. Assuming an intermediate telescope diameter between HabEx and LUVOIR, \citet{2018AJ....155..230G} explored through simulations the impact of false-positives and precursor knowledge on a direct imaging survey yield of Earth-mass analogs. They expanded upon the work in \citet{Stark_2014,Stark_2015}, considering the exoplanet yield of planets besides exo-Earths, and found that 77\% of imaged exoplanets that would at first appear to be exo-Earth analogs at HZ projected orbital separations were in fact other planets in the system at different true orbital separations and planet radii.  They found that precursor knowledge of the orbits would help substantially in reducing this false-positive rate and consequently improving the exoplanet yield.

Estimating the yield of a future flagship direct imaging mission is a well-trodden subject of inquiry as different mission concepts have been proposed over the preceeding decades, dating back to at least \citet{brown2004a,brown2004b,brown2005}. For example, \citet{2007MNRAS.374.1271A} explored detailed analytic estimates of direct imaging mission yield, employing a differential-based formalism of yield estimates, assuming a local stellar density and initial mass function (IMF), a lognormal planet size distribution, and investigating the optimization of yield as a function of observing wavelength, exo-zodiacal (zodi) levels, stellar metallicity, and other considerations, but did not explore the impact of precursor knowledge on survey yield or efficiency. They applied their analytic relations to a suite of mission concepts under consideration at the time. Next, \citet{2010ApJ...715..122B} explored the exo-Earth yield if the design reference mission for the James Webb Space Telescope (JWST) had been equipped with a starshade, employing a sequential observation approach to estimate the probability of observing an exo-Earth with each subsequent observation based upon the outcome of prior observations, deriving an estimate for survey completeness.  \citet{2011PASP..123..171C} expanded upon this work to investigate different observing strategies for detection and confirmation of exo-Earths with a star-shade equipped JWST.  \citet{2012OptEn..51a1002L} employed an analysis investigating the yield for a set of different direct imaging mission aperture sizes for a specific set of the nearest stars, employing numerical yield estimates from a set of analytic scaling dependencies on various mission parameters.  Next, \citet{2013SPIE.8864E..03S} developed a set of numerical simulation for estimating exoplanet yield for a direct imaging mission that could be customized for any mission concept with an end-to-end simulation framework, including applying exoplanet demographics from the \textit{Kepler} mission, and specifically applied this to the Roman mission concept (formerly WFIRST and AFTA) for exoplanets in general, and not specifically exo-Earths. Finally, \citet{kopparapu} used the SAG13 exoplanet demographics from \textit{Kepler} to estimate direct imaging mission yields of different exoplanet types, although those demographics were super-ceded by the exoplanet population demographics in \citet{Dulz} and the final HabEx and LUVOIR yield estimates \citep{gaudi2020habitable,theluvoirteam2019luvoir,Standards}.

Several studies have also looked at the impact on precursor knowledge on direct imaging exoplanet yield. We define precursor knowledge in this work to be knowledge both of which stars have planets of interest for direct imaging, and also sufficient knowledge of projected orbital separation to image the planets outside the inner working angle when targeted. \citet{2016JATIS...2a1020T} explored the direct imaging yield for the Roman mission, where targets were all known prior to imaging, and taking into account specific coronagraphic mask architectures with lab-based sensitivity curves as a function of angular separation. \citet{2010ApJ...720..357S} explored the impact of precursor knowledge from astrometry for a former mission concept called the Occulting Ozone Observatory with a 1.1 m aperture, and in particular identified that the yield of exoplanets could be increased by a factor of 4-5 from precursor knowledge.  However, \citet{2009SPIE.7440E..0BS} conducted numerical simulations of the exo-Earth yield for a future direct imaging mission (in this case the THEIA concept), and found that precursor knowledge from astrometry did not significantly impact mission yield, but did significantly improve direct imaging survey mission efficiency. They also explicitly assess the impact of yield from $\eta_\oplus$, finding that precursor knowledge provides increasing benefits for decreasing exo-Earth occurrence rates. \citet{2011PASP..123..923D} assessed the impact of precursor knowledge from astrometry on the number of re-visits required (survey efficiency) for a set of seven direct imaging targets, finding that precursor knowledge decreases the required number of revisits for a coronagraphic mission and to a lesser extent for an external occulter for a set of four specific prior mission concepts.

In this work, we develop a toy model to derive analytic relations for estimating the exo-Earth yield of a direct imaging mission and its dependence on different mission parameters and specifically the impact of precursor knowledge, relying on a set of a few simplifying assumptions. Our intent is to provide a set of relations derived from our toy model to guide and help validate the more detailed simulations carried out previously for HabEx and LUVOIR, and to be further refined for the Habitable Worlds Observatory in the future.  Our analytic treatment is simpler than in \citet{2007MNRAS.374.1271A}; however, we additionally explore analytic yield dependence on precursor knowledge, complementing the aforementioned works that looked at the impact of precursor knowledge through simulations or specific direct imaging mission architectures.  We also specifically look at the parameters for which a direct imaging survey will be in a ``photon noise limited'' or an ``inner working angle limited'' regime, and the transition between the two.

First, we assume circular orbits, which are common for compact terrestrial planetary systems as found by the Kepler mission as inferred from their mutual inclination and transit duration distributions \citet{Lissauer_2011,Shabram_2016,2012ApJ...761...92F,2014PASP..126...34P}, but larger Jovian planets can more commonly exhibit more eccentric orbits. \citet{Kane2013} in particular explored the impact eccentricity had on whether or not a planet falls outside the inner working angle of a direct imaging survey.  Second, we also assume all stars are single Sun-like stars with identical location HZ orbits with exo-Earths located at 1 au.  In other words, we do not marginalize over distributions in planet radius, insolation flux / orbital distance, nor stellar spectral type. \citet{2011ApJ...733..126C} explored the impact of exoplanet direct imaging yield as a function of stellar spectral type, but primarily for ground-based direct imaging instrumentation.  Third, when we consider the impact of precursor knowledge on mission yield, we assume perfect and complete knowledge -- e.g. we do not consider a scenario in which only a fraction of target stars have precursor knowledge, nor when the orbital knowledge is insufficient to fully predict if an exoplanet is outside an inner working angle, such as can be the case with the radial velocity method with an unknown orbital inclination.  We also do not consider the impact of planet multiplicity on exoplanet yield, where planets at larger orbital separations can mistakenly appear to be projected into Habitable Zone orbits in a single visit \citep{2018AJ....155..230G}.  

Next, we adopt a simplified noise model where the contributions from exo-zodis and speckles (host starlight suppression residuals) scales with the photon noise, and derive a scaling factor by fitting our model to exposure time estimates made with EXOSIMS \citep{Standards}.  Our model reproduces the EXOSIMS exposure times to within 20\% (see ${\S}$\ref{sec:basic:noPrior} and  ${\S}$\ref{sec:HabExLUVOIR}).  While this model effectively ignores Solar System zodiacal light noise contributions, more detailed computational simulations show $>$50\% disagreement amongst themselves for the same target and instrument configuration \citep{Standards}, and is thus an adequate model for the purposes of developing our simplified analytic approach. Finally, we do not model the impact of obscured vs. un-obscured apertures, and segmented vs. single-aperture telescope designs, which has been shown to also introduce important changes in yield as a function of telescope diameter \citep{NAP26141}.

In \hyperref[sec:basicYield]{${\S}$\ref{sec:basicYield}}, we derive a basic yield model for direct imaging surveys, one that is only limited by photon noise (e.g. a negligible inner working angle), without and with precursor knowledge in turn. In \hyperref[sec:iwa]{${\S}$\ref{sec:iwa}}, we enhance that ``photon-noise limited'' model by evaluating the impact of the telescope inner working angle on the random observations of an uninformed survey -- e.g. assessing the fraction of survey time that is lost when a target for which we do not know whether or not the orbital ephemerides lies inside the inner working angle. In \hyperref[sec:iwalim]{${\S}$\ref{sec:iwalim}}, we derive equations to describe the lower bound on the telescope diameter specified by the required yield and the inner working angle, in the ``inner working angle limited'' regime. In \hyperref[sec:limdetect]{${\S}$\ref{sec:limdetect}}, we investigate the transition between the ``photon noise limited'' and ``inner working angle limited'' regimes to assess under what direct imaging survey parameters a given survey would be photon noise or inner working angle limited. In \hyperref[sec:results]{${\S}$\ref{sec:results}}, we summarize our key results to be useful in evaluatng future mission architecture design trades. In \hyperref[sec:discuss]{${\S}$\ref{sec:discuss}}, we compare these analytic relations to more detailed numerical simulations carried out in prior work. In \hyperref[sec:concl]{${\S}$\ref{sec:concl}} we present our conclusions.


\section{Basic Photon Noise Yield Model}\label{sec:basicYield}

In this section, we establish a basic exo-Earth yield model without precursor knowledge and without an inner working angle requirement, and  accounting only for sufficient detected photons from the targeted planets.  In other words, we first assume the planet is always imaged outside the inner working angle, and after we construct this model, we then consider the impacts of precursor knowledge in ${\S}$\ref{sec:basic:prior}, and the impact of inner working angle in ${\S}$\ref{sec:iwa}.

\subsection{No Precursor Knowledge}\label{sec:basic:noPrior}
First, we consider the case of no precursor knowledge: a scenario in which we have no information about the distribution of exo-Earths, and thus target stars are searched at random. We first define as expected: $N_\oplus=N_*\eta_\oplus,$ where $N_*$ is the number of stars we are surveying, and $N_\oplus$ is the number of those stars that host exo-Earth planets. As stated in ${\S}$\ref{intro} for simplification in our analytic model, we assume all exo-Earths are located at an orbital distance equal to 1 au from their host stars, and all host stars are identical and Sun-like, e.g. 1$M_\odot$ and 1$R_\odot$. Further, we assume any information about the insolation flux, size range, or other properties of the exo-Earths are incorporated into the value of $\eta_\oplus$.  In other words, we do not consider a range of insolation flux / habitable zone orbital distance, planet size, or host star spectral type distributions, as explored in \citet{Kane2013,2011ApJ...733..126C}.

Assuming a uniform random distribution of $N_*$ identical stars, we express the stellar density as \begin{equation}\label{eq:rhostar_first}\rho_*= \frac{N_*}{\frac{4}{3}\pi D_{\lim}^3},\end{equation} where $D_{\lim}$ is the limiting distance of our hypothetical survey (due to the assumption of identical stars, . 
We can also express this in terms of the density of exo-Earths as $\rho_\oplus=\rho_*\eta_\oplus,$ so \begin{equation}\label{eq:rhoexpand}\rho_\oplus= \rho_*\eta_\oplus=\frac{N_\oplus}{\frac{4}{3}\pi {D_{\lim}}^3}\end{equation}
Second, we can next define that the total on-sky time $T=\sum_{k=1}^{N_*}t_k$, where $t_k$ is the time spent on the $k$th star and $T$ is the constant survey duration, and where we assume that survey duration is constant, ignoring mission extensions and assuming the mission surveys all $N_*$ stars. Third, we define $R(\nu)$ to be the bolometric rate at which a star isotropically radiates light over the wavelength range of interest, in photons per second, for some central frequency $\nu$. We also assume a constant star-planet flux contrast ratio $K$, e.g. identical Earth-size planets orbiting our assumed and simplistic local universe of identical stars. Then, the rate $R_e$, in photons/sec, at which our survey telescope detects reflected light from the $k$th planet would be \begin{equation}\label{eq:reffective}R_e =RK\frac{\pi (d/2)^2\varepsilon}{4\pi D^2_k}=\frac{RKd^2\varepsilon}{16D^2_k}\end{equation} where $D_k$ is the distance in meters from earth to the $k$th star and its planet, where $d$ is the diameter in meters of the telescope, and where $\varepsilon(f)$ is the telescope efficiency as a function of frequency, including filters, atmospheric interference, etc., and assumed to have negligible throughput degradation over the course of the survey duration. 


Next, most direct imaging missions have some $SNR\geq SNR_0$ requirement in the continuum flux for each planet observed. We assume the bounding scenario where observations achieve the minimum $SNR=SNR_0$ requirement. Reaching that $SNR$ for the $k$th star requires an exposure time \begin{equation*} t_k\approx SNR_0^2\frac{R_e+2B_k}{R_e^2}\end{equation*}
where $B_k$ is the count rate for all sources of background. This is a restatement of the CCD SNR equation solved for the exposure time, with the assumption of zero noise detectors, and combining any scattered light, zodiacal, exo-zodiacal and similar background noise into a single term $B_k$. We next make a simplifying assumption that $B_k$ scales linearly with $R_e$ and thus also the stellar flux $R$, such that $B_k$ can be expressed as $B_k=\frac12 r' R_e$ and where $r'$ can be tuned for a different set of assumed noise levels. This is equivalent to assuming there is no systematic noise floor from Solar System zodiacal light or otherwise, and that independent of the stellar brightness, there is a constant exo-zodiacal contribution for every star that can be combined with the scattered light contribution to the background noise that will scale with the stellar brightness, and that the achieved flux contrast for the direct imaging mission instrument contributes a constant noise term to the planet flux measurement. Further, we ignore for simplicity how this flux contrast varies with angular separation, and assume for example that a `dark hole' \citep{darkholeref} of constant flux contrast $K$ can be created at any location outside an inner working angle for the assumed constant background noise term. Then, we have: \begin{equation*}t_k=SNR_0^2\frac{1+r'}{R_e}.\end{equation*}
To simplify the expression, let $r\equiv1+r'$ and thus $B_k=\frac12 (r-1) R_e$.  Then:

\begin{equation}\label{eq:tkbasic}t_k=\frac{rSNR^2_0}{R_e}=\frac{16 rSNR^2_0 D^2_k}{RKd^2\varepsilon}.\end{equation}

We discuss the validity of adopting this noise model further in ${\S}$\ref{sec:HabExLUVOIR}.
Because the stars are randomly distributed throughout a sphere of radius $D_{\lim},$ we can divide the sphere up into $N_*$ spherical shells of equal volume, and assume that there is exactly one star contained in each spherical shell (e.g. we are ignoring any stellar binarity). Each shell would have volume $\frac{4\pi}{3N_*} D_{\lim}^{3}.$ Therefore, the outer radius of the $k$th spherical shell $D_k$ must satisfy
\begin{equation*}\frac{4\pi}{3}D_{k}^3=\frac{4k\pi}{3N_*}D_{\lim}^3\end{equation*}
\begin{equation*}D_k^3=\frac{k}{N_*}D_{\lim}^3=\frac{3k}{4\pi\rho_* D_{\lim}^3}D_{\lim}^3=\frac{3k}{4\pi\rho_*}\end{equation*}
\begin{equation}\label{eq:tkexpandnoknow}t_k=\frac{16 rSNR^2_0 D^2_k}{RKd^2\varepsilon}=\frac{16 rSNR^2_0 }{RKd^2\varepsilon}\left(\frac{3k}{4\pi\rho_*}\right)^{2/3}\end{equation}
We assume the planet to be located at the outer border of the spherical shell; this is a worst-case scenario for an observer.
Because the total on-sky time for a survey must be equal to the sum of the integration times (ignoring slew time, overheads, etc.),
\begin{equation}\label{eq:Tsumnoknow}T=\frac{16rSNR_0^2}{RKd^2\varepsilon}\left(\frac{3}{4\pi\rho_*}\right)^{2/3}\sum_{k=1}^{N_*} k^{2/3}\end{equation}\newline
So, because $T$ is fixed, we know that
\begin{equation}\label{eq:initd2noknow}d^2=\frac{16rSNR_0^2}{RKT\varepsilon}\left(\frac{3}{4\pi\rho_*}\right)^{2/3}\sum_{k=1}^{N_*} k^{2/3}\end{equation}\newline
\subsubsection{With simple approximation}\label{sec:basic:noPrior:simple}
We approximate \begin{equation}\label{eq:basicapprox}\sum_{k=1}^{N_*} k^{2/3}\approx\int_0^{N_*} x^{2/3} dx=\frac{3}{5}N_*^{5/3}\end{equation} Because the first is effectively a Riemann sum of the second, this is a reasonable approximation. However, it will have a very high percent error for low values of $N_*.$ Therefore, we assume that $N_*$ is large, greater than 100. (A value of $N_*=100$ yields a $0.83\%$ error, and percent error improves with increasing $N_*.$)\newline

So, \begin{equation}d^2=\frac{16rSNR_0^2}{RKT\varepsilon}\left(\frac{3}{4\pi\rho_*}\right)^{2/3}\frac{3}{5}N_*^{5/3}\end{equation}

\begin{equation*}d=4SNR_0\left(\frac{3}{4\pi\rho_*}\right)^{1/3}\sqrt{\frac{3rN_\oplus^{5/3}}{5RKT\varepsilon}\eta_\oplus^{-5/3}}\end{equation*}

\begin{equation}\label{eq:simplenoknowd}d=4SNR_0\eta_\oplus^{-5/6}\sqrt[6]{\frac{243r^3 N_\oplus^{5}}{2000\pi^2\rho_*^2R^3K^3T^3\varepsilon^3}}\end{equation}
We approximate that the cost $c$ of large telescopes scales as $c\propto d^{2.5},$ \citep{10.1117/12.552181}, and then we can say that the cost $c$ can be expressed as \begin{equation*}c=Cd^{2.5}=C\left(4SNR_0\sqrt[6]{\frac{243r^3N_\oplus^5}{2000\pi^2\rho_*^2T^3R^3K^3\varepsilon^3}}\eta_\oplus^{-5/6}\right)^{5/2}\end{equation*} where $C$ is a scaling constant for cost. \begin{equation}c=C\left(\frac{62208\cdot r^3 SNR_0^6\cdot N_\oplus^5}{125\pi^2\rho_*^2T^3R^3K^3\varepsilon^3}\right)^{5/12}\eta_\oplus^{-25/12}\end{equation}\newline
Similar equations can be found for other cost-scaling exponents.
\subsubsection{With advanced approximation}\label{sec:basic:noPrior:adv}
We now use the approximation derived in \nameref{sec:appA}: \begin{equation}\begin{split}\sum_{k=1}^{N_*} k^{2/3} \approx \frac{3(N_*+1)^{5/3}}{5}-\frac{(N_*+1)^{2/3}}{2}-\frac{1}{10}\end{split}\tag{\ref{eq:sumapprox}}\end{equation}
This approximation has far lower error compared to the exact Riemann sum. So, 
\begin{equation*}\begin{split}d^2 = \frac{16rSNR_0^2}{RKT\varepsilon}&\left(\frac{3}{4\pi\rho_*}\right)^{2/3}\cdot\\&\left(\frac{3(N_*+1)^{5/3}}{5}-\frac{(N_*+1)^{2/3}}{2}-\frac{1}{10}\right)\end{split}\end{equation*}
\begin{equation*}\begin{split}d = 4\sqrt{r}SNR_0&\left(\frac{9}{16\pi^2\rho_*^2R^3K^3T^3\varepsilon^3}\right)^{1/6}\cdot\\&\left(\frac{3(N_*+1)^{5/3}}{5}-\frac{(N_*+1)^{2/3}}{2}-\frac{1}{10}\right)^{1/2}\end{split}\end{equation*}
\begin{equation}\begin{split}d=&\left(\frac{2304r^3SNR_0^6}{\pi^2\rho_*^2R^3K^3T^3\varepsilon^3}\right)^{1/6}\cdot\\&\left(\frac{3(N_*+1)^{5/3}}{5}-\frac{(N_*+1)^{2/3}}{2}-\frac{1}{10}\right)^{1/2}\end{split}\end{equation}
Again, we use the approximation that cost scales as $d^{2.5},$ so the cost $c$ is
\begin{equation}\begin{split}c=C&\left(\frac{2304r^3SNR_0^6}{\pi^2\rho_*^2R^3K^3T^3\varepsilon^3}\right)^{5/12}\cdot\\&\left(\frac{3(N_\oplus/\eta_\oplus+1)^{5/3}}{5}-\frac{(N_\oplus/\eta_\oplus+1)^{2/3}}{2}-\frac{1}{10}\right)^{5/4}\end{split}\end{equation}
Again, similar expressions can be found for different exponents for the scaling of cost with telescope diameter.\newline
\newline
\subsubsection{Solving for other variables}
It may be useful to rearrange equation \ref{eq:simplenoknowd} to solve for different variables. A few rearrangements are given here.
\begin{equation}\label{eq:noplus_mostbasic} N_\oplus=\eta_\oplus\sqrt[5]{\frac{125d^6\pi^2\rho_*^2R^3K^3T^3\varepsilon^3}{62208SNR_0^6r^3}}\end{equation}
\begin{equation}T=\sqrt[3]{\frac{62208SNR_0^6r^3N_\oplus^5}{125d^6\pi^2\rho_*^2R^3K^3\varepsilon^3\eta_\oplus^5}}\end{equation}
\subsection{With Precursor Knowledge}\label{sec:basic:prior}
We now evaluate the benefits of precursor knowledge by extending the previously introduced basic yield model. In this section, we assume that the survey has perfect precursor knowledge from previous observations, and only serves to confirm the existence of and characterize the exo-Earths. The lack of consideration for inner working angle in the basic yield model is somewhat more appropriate here, because determination of the orbital ephemerides can be accomplished through the precursor observations; hence we can target the systems with exo-Earths when they are known to be exterior to the telescope's inner working angle. In other words, we assume that over the duration of the survey there is always at least one exo-Earth available to observe outside the inner working angle at any given time for at least one system in the target list, and over the course of the survey duration, all exo-Earth hosts will be targeted when the exo-Earth is exterior to the inner working angle. We revisit this assumption by explicitly considering the impact of an inner working angle in ${\S}$\ref{sec:iwa}. 

Since we now know which stars have target-able exo-Earths, our stellar sample size contains only exo-Earth hosting systems and matches the number of exo-Earths we wish to confirm and characterize, and thus $N_*=N_\oplus$.   However, we aren't changing the density of the stars, only our selection process, so equation \ref{eq:rhoexpand} still applies.
\begin{equation}\rho_\oplus= \rho_*\eta_\oplus=\frac{N_\oplus}{\frac{4}{3}\pi {D_{\lim}}^3}\tag{\ref{eq:rhoexpand}}\end{equation}
Again, the total survey on-sky duration can be expressed as the sum of individual target exposures, again ignoring slew times and overhead:
\begin{equation*}
    \sum\limits_{k=1}^{N_*} t_k = T
\end{equation*}
where $t_k$ is the time spent on the $k$th star, and $T$ is the total on-sky time. Again, the detected photo-electron rate from the $k$th exo-Earth is: 
\begin{equation}R_e =RK\frac{\pi (d/2)^2\varepsilon}{4\pi D^2_k}=\frac{Rd^2\varepsilon}{16D^2_k}\tag{\ref{eq:reffective}}\end{equation}
and the per-target observing time of:
\begin{equation}t_k=\frac{rSNR^2_0}{R_e}=\frac{16 rSNR^2_0 D^2_k}{RKd^2\varepsilon}\tag{\ref{eq:tkbasic}}\end{equation}
We assume again that the stars are randomly distributed throughout a sphere of radius $D_{\lim},$ and thus we can divide the sphere up into $N_\oplus$ spherical shells of equal volume, and assume that there is exactly one star with an exo-Earth contained in each spherical shell. Each shell would have volume $\frac{4\pi}{3N_\oplus} D_{\lim}^{3}.$ Therefore, the outer radius of the $k$th spherical shell $D_k$ must satisfy
\begin{gather}
\frac{4\pi}{3}D_{k}^3=\frac{4k\pi}{3N_\oplus}D_{\lim}^3 \\
D_k^3=\frac{k}{N_\oplus}D_{\lim}^3=\frac{3kN_\oplus}{4\pi\rho_*\eta_\oplus N_\oplus}=\frac{3k}{4\pi\rho_*\eta_\oplus} \\
t_k=\frac{16 rSNR^2_0 D^2_k}{RKd^2\varepsilon}=\frac{16 rSNR^2_0 }{RKd^2\varepsilon}\left(\frac{3k}{4\pi\rho_*\eta_\oplus}\right)^{2/3}\end{gather}
Because the total on-sky time must be equal to the sum of the integration times, again ignoring slew time and other overheads, we have:
\begin{equation}\label{eq:Tsumknow}T=\frac{16rSNR_0^2}{RKd^2\varepsilon}\left(\frac{3}{4\pi\rho_*\eta_\oplus}\right)^{2/3}\sum_{k=1}^{N_\oplus} k^{2/3}\end{equation}\newline
So, because $T$ is fixed, we know that
\begin{equation}\label{eq:initd2know}d^2=\frac{16rSNR_0^2}{RKT\varepsilon}\left(\frac{3}{4\pi\rho_*\eta_\oplus}\right)^{2/3}\sum_{k=1}^{N_\oplus} k^{2/3}\end{equation}\newline
\subsubsection{With simple approximation}\label{sec:basic:prior:simple}

Again, we approximate \begin{equation}\sum_{k=1}^{N_\oplus} k^{2/3}\approx\int_0^{N_\oplus} x^{2/3} dx=\frac{3}{5}N_\oplus^{5/3}.\end{equation} Again, this has a high percent error for small values of $N_\oplus,$ so we assume $N_\oplus\geq100.$ For small surveys, this might be an unreasonable assumption. So, \begin{equation}\label{eq:dsqphotonnoise}d^2=\frac{16rSNR_0^2}{RKT\varepsilon}\left(\frac{3}{4\pi\rho_*\eta_\oplus}\right)^{2/3}\frac{3}{5}N_\oplus^{5/3}\end{equation}

\begin{equation}\label{eq:dphotonnoise}d=4SNR_0\left(\frac{3}{4\pi\rho_*\eta_\oplus}\right)^{1/3}\sqrt{\frac{3rN_\oplus^{5/3}}{5RKT\varepsilon}}\end{equation}
Again, we assume that the cost $c\propto d^{2.5},$ but similar results can be shown for other exponents.
\begin{equation*}c=C\left(4SNR_0\left(\frac{3}{4\pi\rho_*\eta_\oplus}\right)^{1/3}\sqrt{\frac{3rN_\oplus^{5/3}}{5RKT\varepsilon}}\right)^{5/2}\end{equation*}
\begin{equation*}c=C\left(4SNR_0\left(\frac{3}{4\pi\rho_*}\right)^{1/3}\sqrt{\frac{3rN_\oplus^{5/3}}{5RKT\varepsilon}}\eta_\oplus^{-1/3}\right)^{5/2}\end{equation*}
\begin{equation*}c=C\left(4SNR_0\left(\frac{3}{4\pi\rho_*}\right)^{1/3}\sqrt{\frac{3rN_\oplus^{5/3}}{5RKT\varepsilon}}\right)^{5/2}\eta_\oplus^{-5/6}\end{equation*}
\begin{equation}c=C\left(\frac{62208\cdot r^3SNR^6_0N_\oplus^{5}}{125\pi^2\rho^2_* T^3R^3K^3\varepsilon^3}\right)^{5/12}\eta_\oplus^{-5/6}\end{equation}
\newline

\subsubsection{With advanced approximation}\label{sec:basic:prior:adv}
Again, we use the approximation derived in \nameref{sec:appA}: \begin{equation}\sum_{k=1}^{N_\oplus} k^{2/3} \approx \frac{3(N_\oplus+1)^{5/3}}{5}-\frac{(N_\oplus+1)^{2/3}}{2}-\frac{1}{10}\tag{\ref{eq:sumapprox}}\end{equation}
Note that we use $N_\oplus$ instead of $N_*$ to prevent confusion.\newline
So,
\begin{equation*}\begin{split}d^2 = \frac{16rSNR_0^2}{RKT\varepsilon}&\left(\frac{3}{4\pi\rho_*\eta_\oplus}\right)^{2/3}\cdot\\&\left(\frac{3(N_\oplus+1)^{5/3}}{5}-\frac{(N_\oplus+1)^{2/3}}{2}-\frac{1}{10}\right)\end{split}\end{equation*}
\begin{equation*}\begin{split}d = 4\sqrt{r}SNR_0&\left(\frac{9}{16\pi^2\rho_*^2\eta_\oplus^2R^3K^3T^3\varepsilon^3}\right)^{1/6}\cdot\\&\left(\frac{3(N_\oplus+1)^{5/3}}{5}-\frac{(N_\oplus+1)^{2/3}}{2}-\frac{1}{10}\right)^{1/2}\end{split}\end{equation*}
\begin{equation}\begin{split}d = &\left(\frac{2304r^3SNR_0^6}{\pi^2\rho_*^2R^3K^3T^3\varepsilon^3}\right)^{1/6}\cdot\\&\left(\frac{3(N_\oplus+1)^{5/3}}{5}-\frac{(N_\oplus+1)^{2/3}}{2}-\frac{1}{10}\right)^{1/2}\eta_\oplus^{-1/3}\end{split}\end{equation}
Again, we will be making the assumption that cost scales as $d^{2.5},$ so the cost $c$ is

\begin{equation}\begin{split}c = C&\left(\frac{2304r^3SNR_0^6}{\pi^2\rho_*^2R^3K^3T^3\varepsilon^3}\right)^{5/12}\cdot\\&\left(\frac{3(N_\oplus+1)^{5/3}}{5}-\frac{(N_\oplus+1)^{2/3}}{2}-\frac{1}{10}\right)^{5/4}\eta_\oplus^{-5/6}\end{split}\end{equation}

Again, similar expressions can be found for different exponents.\newline

\subsubsection{Solving for other variables}
To solve for $N_\oplus,$ we begin with equation \ref{eq:dsqphotonnoise}:
\begin{equation}d^2=\frac{16rSNR_0^2}{RKT\varepsilon}\left(\frac{3}{4\pi\rho_*\eta_\oplus}\right)^{2/3}\frac{3}{5}N_\oplus^{5/3}\tag{\ref{eq:dsqphotonnoise}}\end{equation}
We rearrange the terms:
\begin{equation}\label{eq:noplusprior}
    N_\oplus=\sqrt[5]{\frac{125\pi^2 d^6 R^3 K^3 T^3 \varepsilon^3 \rho_*^2 \eta_\oplus^2}{62208 r^3 SNR_0^6}}
\end{equation}
\newline\newline
To solve for $SNR_0$ in terms of the other variables, we begin with equation \ref{eq:dphotonnoise}:
\begin{equation}d=4SNR_0\left(\frac{3}{4\pi\rho_*\eta_\oplus}\right)^{1/3}\sqrt{\frac{3rN_\oplus^{5/3}}{5RKT\varepsilon}}\tag{\ref{eq:dphotonnoise}}\end{equation}
We rearrange the equation and simplify.
\begin{equation}\label{eq:SNR0prior}SNR_0=\sqrt[3]{\frac{4\pi\rho_*\eta_\oplus}{3}}\sqrt{\frac{5RKT\varepsilon d^2}{48rN_\oplus^{5/3}}}\end{equation}

To solve for $T$, we begin with equation \ref{eq:Tsumknow}.
\begin{equation}T=\frac{16rSNR_0^2}{RKd^2\varepsilon}\left(\frac{3}{4\pi\rho_*\eta_\oplus}\right)^{2/3}\sum_{k=1}^{N_\oplus} k^{2/3}\tag{\ref{eq:Tsumknow}}\end{equation}
We again approximate $\sum_{k=1}^{N_\oplus}\approx\frac35 N_\oplus^{5/3}:$
\begin{equation}\label{eq:Tsumknow_sub}T=\frac{48rSNR_0^2 N_\oplus^{5/3}}{5RKd^2\varepsilon}\left(\frac{3}{4\pi\rho_*\eta_\oplus}\right)^{2/3} \end{equation}

\section{Photon Noise Yield Model Accounting for IWA}\label{sec:iwa}

Now we introduce a more complicated model, in which we account for how the inner working angle impacts the target exposure times. In order to model the impact of the inner working angle on yield, one might scale the required on-sky time for each exo-Earth inversely by the ``time fraction usable,'' the percentage of time that the exo-Earth spends outside of the telescope's inner working angle, as a function of that each individual exo-Earth's inclination. Such a case-by-case scaling for individual exo-Earth inclinations would not yield a simple analytic approximation for the total survey duration. Instead, we average the time fraction usable of the exo-Earth over all inclinations assuming uniform random distribution in the cosine of the inclination. Then, we scale the total on-sky time inversely by that average, to obtain an approximation of the time needed to achieve the required yield in an uninformed survey.  For some targets, with face-on inclinations the required time will be shorter than average (a larger fraction of the time the target will be outside the $iwa$), whereas for edge-on targets the required time will be longer than average (a smaller fraction of the time the target will be outside the $iwa$).

We make some additional simplifications and assumptions.  Again, we do not account for variable spectral types, or distances of the exo-Earths from their host stars -- we assume all exo-Earths orbit at 1 au from a Sun-like star.  We also do not account for revisits in our analysis at different orbital phases.  We also assume that only a survey with precursor knowledge can wait for the right time to target a given exo-Earth when it is outside the $iwa$, whereas an uninformed survey will sometimes observe a system when the exo-Earth is inside the $iwa$.

Finally, while we continue to assume that the target stars are uniformly distributed in a spherical volume with radius $D_{\lim}$ in calculating exposure times, for assessing the impact of the $iwa$, we instead assume that all planets are at $D_{\lim}$, a worst case scenario in assessing the fraction of time a given exo-Earth is external to the $iwa$. Scaling this $iwa$ impact with $D_k$ instead of $D_{\lim}$ does not yield an analytic sum for the survey duration, although this can be computed numerically, which we next show.

\subsection{No Precursor Knowledge}\label{sec:iwa:noPrior}
An uninformed survey will be forced to target potential exo-Earths randomly in orbital phase, without any initial knowledge of their orbital ephemerides. Thus, the efficiency of observations would be proportional to the average percentage of time in which the planet is observable. Because we are considering the impact of the inner working angle, we assume that the planet is observable when outside the telescope's inner working angle, and unobservable other times.  Note, the same is not true for revisits, which are not considered herein, but will asymptote to the precursor knowledge case as the orbital ephemerides are constrained and thus observations can be optimally timed after the initial detection and with improvements in orbital determination. 

As before, $T=\sum^{N_*}_{k=1}t_k.$ Because some fraction of that time $t_k$ is unusable (e.g. when the planet is inside the $\text{iwa}$), we can express it as $t_k=u_k+w_k,$ where $u_k$ is the usable time, and $w_k$ is the unusable time. We have an expression for how much usable time we need: \begin{equation}u_k=\frac{16 rSNR^2_0 }{RKd^2\varepsilon}\left(\frac{3k}{4\pi\rho_*}\right)^{2/3}\tag{\ref{eq:tkexpandnoknow}}\end{equation}
As derived in \nameref{sec:appB}, the fraction of time usable for a given exoplanet can be expressed as the piecewise function
\begin{equation*}t_f=
\begin{dcases}
0&s_c\geq a\\
\frac{2}{\pi}\cdot\arccos\left(\frac1a\cdot\sqrt{\frac{\text{iwa}^2-a^2 \cos^2i} {1-(\cos^2i)}} \right)
&a\cos{i}<s_c<a\\
1&a\cos i>s_c
\end{dcases}\end{equation*}
where $a$ is the exo-Earth's semi-major axis, $s_c = D_{k}\cdot\text{iwa}$ is the projection of the inner working angle at the distance to the target star (we assume the sphere centered on the observer and intersecting the star to be tangentially flat, such that a flat projection may be assumed), and $\cos i$ is the cosine of the inclination. \newline
Because $\cos i$ is uniform random, the average time fraction usable can be obtained by integrating the expression above $d\cos i$ from $0$ to $1.$ That yields the equation
\begin{equation}t_{a}=\frac{\sqrt{a^2-s_c^2}}{a}\tag{\ref{eq:tfracavg}}\end{equation}
where $t_a$ is the average time fraction usable.\newline
We can say that, on average
\begin{equation}
    t_k = \frac{u_k}{t_a} = \frac{a}{\sqrt{a^2-s_c^2}}\cdot\frac{16 rSNR^2_0}{RKd^2\varepsilon}\left(\frac{3k}{4\pi\rho_*}\right)^{2/3}\label{eq:iwa_noprior_unsimplified}
\end{equation}

We note that the average time spent observing a target for which an exoplanet is outside the \text{iwa} is a simplifying assumption - e.g. that we are uniform randomly observing this target in time, as opposed to observing this target with a cadence that maximizes the probability that the planet is captured outside the \text{iwa}. This is thus a bounding worst-case scenario.

To solve for $d$, we recall that $s_c=D_{k}\cdot\text{iwa}.$  However, substituting this in Equation $\ref{eq:iwa_noprior_unsimplified}$ and summing over all the targets results in a non-trivial summation in k:
\begin{equation}
    T=\frac{16rSNR_0^2}{RKd^2\varepsilon}\left(\frac{3}{4\pi\rho_*}\right)^{1/3}\sum^{N_*}_{k=1}\frac{k^{1/3}}{\sqrt{\left(\frac{4\pi\rho_*}{3k}\right)^{2/3} - \left(\frac{\text{iwa}}{a}\right)^2}}
\end{equation}
Instead, we assume a worst-case scenario -- in correcting for the average fraction of the time that the planet is external to the inner working angle, we assume that all the exo-Earths are at the survey limiting distance $D_{\lim}$ and thus $s_c = D_{\lim}\cdot\text{iwa}$ is independent of $k$. Alternatively, we could have taken the average distance, $<D_k>\: =\: 2^{-1/3} D_{\lim} \sim 0.79 \:D_{\lim}$. Next, we approximate \citep{Mawet_2012} \begin{equation}\label{eq:iwaApprox}\text{iwa}\approx\frac{n_i\lambda}{d}\text{ rad}\end{equation} for some $n_i\approx3.$ Then we have from Equation \ref{eq:iwa_noprior_unsimplified}:
\begin{equation*}
    T=\frac{16rSNR_0^2a}{RKd^2\varepsilon\sqrt{a^2-s_c^2}}\left(\frac{3}{4\pi\rho_*}\right)^{2/3}\sum_{k=1}^{N_*} k^{2/3}
\end{equation*}
\begin{equation}
    T=\frac{16rSNR_0^2}{RKd^2\varepsilon\sqrt{1-\left(\frac{s_c}{a}\right)^2}}\left(\frac{3}{4\pi\rho_*}\right)^{2/3}\sum_{k=1}^{N_*} k^{2/3}
\end{equation}
\begin{equation}T=\frac{16rSNR_0^2}{RK\varepsilon d^2\sqrt{1-\left(\frac{n_i D_{\lim}\lambda}{ad}\right)^2}}\left(\frac{3}{4\pi\rho_*}\right)^{2/3}\sum_{k=1}^{N_*} k^{2/3}\label{eq:Tnoprioriwa}\end{equation}
which has a more trivial summation in k and is independent of $\eta_\oplus$.  From the above, $T$ diverges as $d$ approaches $\frac{3D_{\lim}\lambda}{a}$. This is as expected -- planets with semi-major axes approaching the inner working angle have a fraction of time observable outside the inner working angle that limits to zero. We discuss the impact of this in \ref{sec:limdetect:noprior}.

Next, for simplification of presentation, we define:
\begin{align}n&\equiv \frac{n_i\lambda}{a}\label{eq:ndef}\\
m&\equiv\frac{16rSNR_0^2}{RK\varepsilon}\left(\frac{3}{4\pi\rho_*}\right)^{2/3}\label{eq:mdef}\end{align}
Where $\lambda$ is the wavelength at which we are observing. Note that neither $m$ nor $n$ has any dependence on $\eta_\oplus.$
Then, \begin{equation}\label{eq:Tnoprioriwa_simplified}T=\frac{m}{ d^2\sqrt{1-\left(\frac{D_{\lim}n}{d}\right)^2}}\sum_{k=1}^{N_*} k^{2/3}\end{equation}
The solution to this, as derived in \nameref{sec:appC}, is
\begin{equation}d=\pm\sqrt{\frac{D_{\lim}^2 n^2}{2}\pm\frac{1}{2}\sqrt{D_{\lim}^4 n^4+\frac{4m^2 \left(\sum_{k=1}^{N_*} k^{2/3}\right)^2}{T^2}}}\end{equation}
We can remove some common factors:
\begin{equation*}d=\pm\frac{D_{\lim}n}{\sqrt{2}}\sqrt{1\pm\sqrt{1+\frac{4m^2\left(\sum_{k=1}^{N_*}k^{2/3}\right)^2}{D_{\lim}^4 n^4 T^2}}}
\end{equation*}
The diameter can't be negative, so we can eliminate the negative solutions:
\begin{equation*}d=\frac{D_{\lim}n}{\sqrt{2}}\sqrt{1\pm\sqrt{1+\frac{4m^2\left(\sum_{k=1}^{N_*}k^{2/3}\right)^2}{D_{\lim}^4 n^4 T^2}}}
\end{equation*}
Also,  the inner square root contains $1$ plus some non-negative number, so the result of the inner square root is at least $1.$ Evaluating the $-$ of the $\pm$ for the outer square root would require taking the square root of a negative number, which would result in a complex diameter. The telescope diameter must be a real number, so we can eliminate the $-$ case.  Thus,
\begin{equation}\label{eq:dsimpleiwa}d=\frac{D_{\lim}n}{\sqrt{2}}\sqrt{1+\sqrt{1+\frac{4m^2\left(\sum_{k=1}^{N_*}k^{2/3}\right)^2}{D_{\lim}^4 n^4 T^2}}}
\end{equation}
\subsubsection{With simple Approximation}\label{sec:iwa:noPrior:simple}
Again, we assume that
\begin{equation}\sum_{k=1}^{N_*} k^{2/3}\approx\int_0^{N_*} x^{2/3} dx=\frac{3}{5}N_*^{5/3}\tag{\ref{eq:basicapprox}}\end{equation}
Again, this has a percent error $>$0.83\% for values of $N_*<100$, so we assume $N_*\geq100.$
Substituting that in, \begin{equation*}d = \frac{D_{\lim}n}{\sqrt{2}}\sqrt{1+\sqrt{1+\frac{4m^2\left(\frac{3}{5}N_*^{5/3}\right)^2}{D_{\lim}^4 n^4 T^2}}}\end{equation*}
\begin{equation}\label{eq:dcomplicatedsmall}d = \frac{D_{\lim}n}{\sqrt{2}}\sqrt{1+\sqrt{1+\frac{36m^2N_*^{10/3}}{25D_{\lim}^4 n^4 T^2}}}\end{equation}
Because \begin{equation}N_*=\frac{4\pi\rho_* D_{\lim}^3}{3},\tag{\ref{eq:rhostar_first}}\end{equation}
we can split up the $N_*^{10/3}$ into $N_*^2\cdot N_*^{4/3},$ and cancel out the factors of $D_{\lim}$:
\begin{equation*}d = \frac{D_{\lim}n}{\sqrt{2}}\sqrt{1+\sqrt{1+\frac{36m^2N_*^{2}}{25 n^4 T^2}\cdot\frac{N_*^{4/3}}{D_{\lim}^4}}}\end{equation*}
\begin{equation}d = \frac{D_{\lim}n}{\sqrt{2}}\sqrt{1+\sqrt{1+\frac{36m^2N_*^{2}}{25 n^4 T^2}\left(\frac{4\pi\rho_*}{3}\right)^{4/3}}}\end{equation}
To substitute back in for our simplifying variable $m^2,$ we square equation \ref{eq:mdef}:
\begin{equation*}m^2=\frac{256r^2SNR_0^4}{R^2K^2\varepsilon^2}\left(\frac{3}{4\pi\rho_*}\right)^{4/3}\end{equation*}
to get:
\begin{equation*}d = \frac{D_{\lim}n}{\sqrt{2}}\sqrt{1+\sqrt{1+\frac{36\cdot256r^2SNR_0^4 N_*^{2}}{25 R^2K^2\varepsilon^2 n^4 T^2}\left(\frac{3\cdot4\pi\rho_*}{3\cdot4\pi\rho_*}\right)^{4/3}}}\end{equation*}
which simplifies to:
\begin{equation}\label{eq:dphotonnoprioriwa}d = \frac{D_{\lim}n}{\sqrt{2}}\sqrt{1+\sqrt{1+\frac{9216 r^2SNR_0^4 N_*^{2}}{25 R^2K^2\varepsilon^2 n^4 T^2}}}\end{equation}
To find the $\eta_\oplus$ dependence, we can expand $N_*$ and $D_{\lim}:$
\begin{equation}\label{eq:dphotonnoprioriwaeta}d = \frac{n}{\sqrt{2}}\sqrt[3]{\frac{3N_\oplus}{4\pi\rho_*\eta_\oplus}}\sqrt{1+\sqrt{1+\frac{9216 r^2SNR_0^4 N_{\oplus}^{2}}{25 R^2K^2\varepsilon^2 n^4 T^2\eta_\oplus^2}}}\end{equation}
To find the cost, we again assume $c=Cd^{2.5}:$
\begin{equation}c = C\left(\frac{n}{\sqrt{2}}\sqrt[3]{\frac{3N_\oplus}{4\pi\rho_*\eta_\oplus}}\sqrt{1+\sqrt{1+\frac{9216 r^2SNR_0^4 N_{\oplus}^{2}}{25 R^2K^2\varepsilon^2 n^4 T^2\eta_\oplus^2}}}\right)^{5/2}\end{equation}
where \begin{equation}n=\frac{n_i\lambda}{a}\tag{\ref{eq:ndef}}\end{equation}

As before, a better approximation can be derived using a more accurate finite summation for $k^{2/3}$, as done in earlier sections, which we do not explicitly carry out herein.

\subsubsection{Solving for other variables}
To solve for $N_\oplus,$ we begin with equation \ref{eq:Tnoprioriwa_simplified}:
\begin{equation}T=\frac{m}{ d^2\sqrt{1-\left(\frac{D_{\lim}n}{d}\right)^2}}\sum_{k=1}^{N_*} k^{2/3}\tag{\ref{eq:Tnoprioriwa_simplified}}\end{equation}
We substitute the definition for $D_{\lim}$ found in equation \ref{eq:rhostar_first}, and again approximate $\sum_{k=1}^{N_*} k^{2/3}\approx\frac35 N_*^{5/3}.$
\begin{equation}\label{eq:Tnoprioriwa_sub}T\approx\frac{3mN_*^{5/3}}{5d^2\sqrt{1-\left(\frac{3N_* n^3}{4\pi\rho_* d^3}\right)^{2/3}}}\end{equation}
The only term with any $N_\oplus$ dependence is $N_*=N_\oplus/\eta_\oplus.$ 
To simplify the presentation of our solution, we define:
\begin{equation*}\alpha\equiv\frac{5Td^2}{3m}\end{equation*}
\begin{equation*}\beta\equiv\frac{n^2}{d^2}\left(\frac{3}{4\pi\rho_*}\right)^{2/3}\end{equation*}
Then, the above equation could be rewritten as
\begin{equation}\label{eq:nstarnoprioriwasolve}\alpha=\frac{N_*^{5/3}}{\sqrt{1-\beta N_*^{2/3}}}\end{equation}
We square to eliminate the square root, and rearrange.
\begin{equation*}\alpha^2=\frac{N_*^{10/3}}{1-\beta N_*^{2/3}}\end{equation*}
\begin{equation*}
    \alpha^2 - \alpha^2\beta N_*^{2/3}=N_*^{10/3}
\end{equation*}
\begin{equation}\label{eq:Nstarpolynomial}
    N_*^{10/3}+ \alpha^2\beta N_*^{2/3} - \alpha^2=0
\end{equation}
$N_*$ is the positive real root of this polynomial which can be computed numerically, and $N_\oplus=\eta_\oplus N_*$ may be derived from $N_*.$ 
\newline\newline
To solve for $SNR_0,$ we begin with equation \ref{eq:Tnoprioriwa}:
\begin{equation}T=\frac{16rSNR_0^2}{RK\varepsilon d^2\sqrt{1-\left(\frac{n_i D_{\lim}\lambda}{ad}\right)^2}}\left(\frac{3}{4\pi\rho_*}\right)^{2/3}\sum_{k=1}^{N_*} k^{2/3}\tag{\ref{eq:Tnoprioriwa}}\end{equation}
We approximate $\sum_{k=1}^{N_*} k^{2/3}\approx\frac35 N_*^{5/3}:$
\begin{equation*}T=\frac{48rSNR_0^2 N_*^{5/3}}{5RK\varepsilon d^2\sqrt{1-\left(\frac{n_i D_{\lim}\lambda}{ad}\right)^2}}\left(\frac{3}{4\pi\rho_*}\right)^{2/3}\end{equation*}
We rearrange the equation.
\begin{equation*}SNR_0^2=\frac{5RKT\varepsilon d^2\sqrt{1-\left(\frac{n_i D_{\lim}\lambda}{ad}\right)^2}}{48rN_*^{5/3}}\left(\frac{4\pi\rho_*}{3}\right)^{2/3}\end{equation*}
And we take a square root.
\begin{equation}\label{eq:SNR0noprior_iwa}SNR_0=\sqrt{\frac{5RKT\varepsilon d^2\sqrt{1-\left(\frac{n_i D_{\lim}\lambda}{ad}\right)^2}}{48rN_*^{5/3}}}\sqrt[3]{\frac{4\pi\rho_*}{3}}\end{equation}

\subsection{Precursor Knowledge}\label{sec:iwa:prior}
Next, we assume we have precursor knowledge of the stars that possess Earth-sized exoplanets in their Habitable Zones, and precursor knowledge of their orbital ephemerides (e.g. period and orbital phase), modulo an unknown inclination. This is thus a bounding best-case scenario. With an unknown inclination, there is still an unknown fraction of time that the exo-Earth is inside the \text{iwa}, and thus the planet may not be observable at all phases (unless $a\cos i>s_c$). However, the planet can be observed at quadrature.  The planet will always be outside the \text{iwa} at quadrature, provided the telescope diameter is adequate (e.g. $a>s_c$). Thus this scenario reduces to assessing the number of stars for which the condition $a>s_c$ is satisfied.  

We can assume that we can target a given exoplanet host star at a time such that the exo-Earth is guaranteed to be observable and located exterior to the observatory \text{iwa}, modulo the unknown inclination. We will also assume for simplicity that mission observations can be scheduled so that each exoplanet is targeted at quadrature, and that there is always a planet available to target at quadrature.  Further, if there are two planets at quadrature at the same over-lapping time, we assume one planet's observations can be deferred to a subsequent orbit without extending the mission lifetime.

Observing the exo-Earth at quadrature will not constrain the inclination of the exo-Earth orbit significantly, and thus an additional observation will be required to constrain the inclination.  Thus we can assume a minimum of two visits per planet would be required in this scenario, but this would also be true of all the other scenarios considered herein.

Because of the targeted observations, no time will be lost due to bad timing (observing when the planet is inside the $iwa$. Consequently, as long as photon noise is the limiting factor for the telescope diameter, it should be the same as was derived in section ${\S}$\ref{sec:basic:prior}.

Using the simpler approximation, 
\begin{equation*}c=C\left(\frac{62208\cdot r^3SNR^6_0N_\oplus^{5}}{125\pi^2\rho^2_* T^3R^3K^3\varepsilon^3}\right)^{5/12}\eta_\oplus^{-5/6}\end{equation*}

Note that the equation for total on-sky time is the same as before:
\begin{equation}\label{eq:T_prior_iwa}
    T=\frac{48rSNR_0^2 N_\oplus^{5/3}}{5RKd^2\varepsilon}\left(\frac{3}{4\pi\rho_*\eta_\oplus}\right)^{2/3}
\end{equation}

This establishes an upper bound on the survey yield, for a given total on-sky time:
\begin{equation*}
    N_\oplus^{5/3} = \left(\frac{4\pi\rho_*\eta_\oplus}{3}\right)^{2/3}\frac{5RKTd^2\varepsilon}{48rSNR_0^2}
\end{equation*}
\begin{equation*}
    N_\oplus^{5} = \left(\frac{4\pi\rho_*\eta_\oplus}{3}\right)^{2}\frac{125R^3K^3T^3d^6\varepsilon^3}{110592r^3SNR_0^6}
\end{equation*}
\begin{equation}\label{eq:noplus_prior_iwa}
    N_\oplus = \sqrt[5]{\frac{125\pi^2\rho_*^2\eta_\oplus^2 R^3 K^3 T^3 d^6\varepsilon^3}{62208r^3SNR_0^6}}
\end{equation}
Similarly, we can establish a minimum telescope diameter.
\begin{equation}\label{eq:d_prior_iwa}
    d=\left(\frac{3}{4\pi\rho_*\eta_\oplus}\right)^{1/3}\sqrt{\frac{48rSNR_0^2 N_\oplus^{5/3}}{5RKT\varepsilon}}
\end{equation}
Similar expressions can be found using the more advanced approximation derived in \nameref{sec:appA}:
\begin{equation*}\begin{split}c = C&\left(\frac{2304r^3SNR_0^6}{\pi^2\rho_*^2R^3K^3T^3\varepsilon^3}\right)^{5/12}\cdot\\&\left(\frac{3(N_\oplus+1)^{5/3}}{5}-\frac{(N_\oplus+1)^{2/3}}{2}-\frac{1}{10}\right)^{5/4}\eta_\oplus^{-5/6}\end{split}\end{equation*}

\section{Inner Working Angle Limited Yield Model}\label{sec:iwalim}
The requirement that all Exo-Earths must have a projected semi-major axis greater than the telescope's projected inner working angle ($a>s_c$) creates a hard lower bound for the telescope diameter, which is referred to as the ``$iwa$ limited'' regime. In some situations, this lower bound is greater than the diameter otherwise required by photon noise limited regime as we have explored in ${\S}$\ref{sec:basicYield} and ${\S}$\ref{sec:iwa}. In this section, we seek to derive an expression for the telescope diameter in such an $iwa$ limited scenario. Note that this expression is applicable to both precursor and no-precursor knowledge cases; precursor knowledge has no impact on this requirement.

To evaluate the requirement that $a>s_c$ for all of our target stars, we examine the definition that we set out in \nameref{sec:appB}:
\begin{equation}s_c=D_{\lim}\cdot\text{iwa}\tag{\ref{eq:scdef}}\end{equation}
In order to satisfy our requirement, this must be less than $a,$ the semi-major axis of the exo-Earth, determined by the position of the habitable zone for that star. Because we are primarily considering solar analogues in our toy model, $a\sim1$ au.\newline
Again, we approximate \begin{equation}\text{iwa}\approx\frac{n_i\lambda}{d} \text{rad},\tag{\ref{eq:iwaApprox}}\end{equation}
where $\lambda$ is the observational wavelength.\newline
So, \begin{equation}\label{eq:scsub}s_c=\frac{n_i D_{\lim}\lambda}{d}.\end{equation}
In order to satisfy the condition, $s_c$ must be at least \begin{equation*}s_c = \frac{n_i D_{\lim}\lambda}{d} = a.\end{equation*} 
Solving for $d$, we get
\begin{equation}\label{eq:dminiwa}d = \frac{n_i D_{\lim}\lambda}{a}.\end{equation}
Substituting $D_{\lim}$ in terms of $\eta_\oplus:$
\begin{equation}\label{eq:dminiwaeta}d = \frac{n_i\lambda}{a}\sqrt[3]{\frac{3N_\oplus}{4\pi\rho_*\eta_\oplus}} = \frac{n_i\lambda}{a}\sqrt[3]{\frac{3N_\oplus}{4\pi\rho_*}}\eta_\oplus^{-1/3}\end{equation}
This is the minimum value for $d,$ based on the inner working angle.
Note that this expression applies to situations with and without precursor knowledge.

Again assuming that the cost $c=Cd^{2.5},$
\begin{equation}c=C\left(\frac{n_i\lambda}{a}\sqrt{\frac{3N_\oplus}{4\pi\rho_*}}\right)^{2.5}\eta_\oplus^{-5/6}\end{equation}

\subsection{No Precursor Knowledge}\label{sec:iwalim:noprior}
If we assume the telescope diameter is defined by Equation \ref{eq:dminiwaeta} in the inner working angle limited regime, we next derive the total on-sky time required by this survey. Only our diameter has changed, so the target scheduling is still the same. The equation for $T,$ then, is the same as before in ${\S}$\ref{sec:iwa:noPrior}:
\begin{equation}T=\frac{16rSNR_0^2}{RK\varepsilon d^2\sqrt{1-\left(\frac{n_i D_{\lim}\lambda}{ad}\right)^2}}\left(\frac{3}{4\pi\rho_*}\right)^{2/3}\sum_{k=1}^{N_*} k^{2/3}\tag{\ref{eq:Tnoprioriwa}}\end{equation}
Again, we have that the survey duration diverges as $d\rightarrow \frac{n_i D_{\lim}\lambda}{a}$ and the survey duration is a worst-case scenario under the assumption that all planets are located at $D_k \rightarrow D_{\lim}$ for estimating the fraction of survey time a given target is located outside the $iwa$.

\subsection{Precursor Knowledge}\label{sec:iwalim:prior}

Since only our expression for the minimum telescope diameter has changed, our total on-sky time is the same:
\begin{equation}T=\frac{16rSNR_0^2}{RKd^2\varepsilon}\left(\frac{3}{4\pi\rho_*\eta_\oplus}\right)^{2/3}\sum_{k=1}^{N_\oplus} k^{2/3}\tag{\ref{eq:Tsumknow}} \end{equation}
Substituting $d=\frac{n_i D_{\lim}\lambda}{a}$, the minimum telescope diameter required for the most distant target to have a planet external to the $iwa$:
\begin{equation}T=\frac{16rSNR_0^2a^2}{n_i^2RK\varepsilon D^2_{\lim}\lambda^2}\left(\frac{3}{4\pi\rho_*\eta_\oplus}\right)^{2/3}\sum_{k=1}^{N_\oplus} k^{2/3}\end{equation}
We use the simple approximation $\sum_{k=1}^{N_\oplus} k^{2/3}\approx \frac35 N_\oplus^{5/3}$: 


\begin{equation}T=\frac{48rSNR_0^2a^2N_\oplus^{5/3}}{5n_i^2RK\varepsilon D^2_{\lim}\lambda^2}\left(\frac{3}{4\pi\rho_*\eta_\oplus}\right)^{2/3}\end{equation}
We substitute the definition of $D_{\lim}$ found by rearranging equation \ref{eq:rhoexpand} and simplifying:

\begin{equation}T=\frac{48rSNR_0^2a^2N_\oplus}{5n_i^2RK\varepsilon \lambda^2}\end{equation}
which has a linear dependence with $N_\oplus$.

\subsubsection{Solving for other variables}
We rearrange equation \ref{eq:dminiwaeta} to solve for $N_\oplus$ and $\eta_\oplus$. The maximum $N_\oplus$ for a given set of parameters that satisfies $a>s_c$ is:
\begin{equation}\label{eq:noplusmaxiwa}N_\oplus < \frac{4\pi\rho_*\eta_\oplus a^3d^3}{3n_i^3\lambda^3}\end{equation}
and the minimum necessary $\eta_\oplus$ for a given set of parameters that satisfies $a>s_c$ is:
\begin{equation}\label{eq:etaminiwa}\eta_\oplus > \frac{3n_i^3\lambda^3 N_\oplus}{4\pi\rho_* a^3d^3}\end{equation}

\section{When are we $iwa$ limited versus photon noise limited in our telescope diameter?}\label{sec:limdetect}

Up until this point, we have estimated the required minimum telescope diameters and survey durations for a set of simplified direct imaging mission parameters, first considering when we are limited by photon noise in ${\S}$\ref{sec:basicYield} and ${\S}$\ref{sec:iwa}, and second when we are limited by inner working angle in ${\S}$\ref{sec:iwalim}, both with and without precursor knowledge.  We now derive the transition between these two regimes to arrive at a prescription to determine when and under what combination of assumed mission parameters a direct imaging survey is photon noise or inner working angle limited.

\subsection{No Precursor Knowledge}\label{sec:limdetect:noprior}
When we are limited by photon noise, including time lost due to unlucky timing when a planet is inside the $iwa$, the necessary diameter is 
\begin{equation}d_\text{noise} = \frac{D_{\lim}n}{\sqrt{2}}\sqrt{1+\sqrt{1+\frac{36m^2N_*^{10/3}}{25D_{\lim}^4 n^4 T^2}}}\tag{\ref{eq:dcomplicatedsmall}}\end{equation}
When we are limited by the inner working angle, 
\begin{equation}d_\text{iwa}=\frac{n_i D_{\lim}\lambda}{a}\tag{\ref{eq:dminiwa}}\end{equation}
At the intersection, they must be equal.
\begin{equation*}\frac{n_i D_{\lim}\lambda}{a}=\frac{D_{\lim}n}{\sqrt{2}}\sqrt{1+\sqrt{1+\frac{36m^2N_*^{10/3}}{25D_{\lim}^4 n^4 T^2}}}\end{equation*}
Substituting the first $n=\frac{n_i \lambda}{a}:$
\begin{equation*}\frac{n_i D_{\lim}\lambda}{a}=\frac{n_i D_{\lim}\lambda}{a\sqrt{2}}\sqrt{1+\sqrt{1+\frac{36m^2N_*^{10/3}}{25D_{\lim}^4 n^4 T^2}}}\end{equation*}
We cancel the common factors and multiply by $\sqrt{2}:$
\begin{equation*}\sqrt{2}=\sqrt{1+\sqrt{1+\frac{36m^2N_*^{10/3}}{25D_{\lim}^4 n^4 T^2}}}\end{equation*}

This results in the trivial expression that:
\begin{equation*}0=\frac{36m^2N_*^{10/3}}{25D_{\lim}^4 n^4 T^2}\end{equation*}
Recall that 
\begin{equation}N_*=\frac{4\pi\rho_*D_{\lim}^3}{3}\tag{\ref{eq:rhostar_first}}\end{equation}
Again, we split up $N_*^{10/3}$ into $N_*^2\cdot N_*^{4/3},$ and cancel the factors of $D_{\lim}:$
\begin{equation*}\frac{36m^2N_*^2}{25n^4T^2}\left(\frac{4\pi\rho_*}{3}\right)^{4/3}=0\end{equation*}
Again, we substitute in for $m^2$ and $n^4$, take the square root and simplify:

\begin{equation}\label{eq:endingzerointersection}\frac{rSNR_0^2N_\oplus a^2}{RK\varepsilon Tn_i^2\lambda^2\eta_\oplus}=0\end{equation}

This condition is met under a few possible trivial scenarios:

\begin{equation*}\begin{cases}
SNR_0=0 & \parbox[t]{5cm}{We don't have to collect any photons.}\\
N_\oplus=0 & \parbox[t]{5cm}{The survey yield is zero.}\\
a=0 & \parbox[t]{5cm}{The targets are impossible to observe.}\\
RK\varepsilon T=\infty & \parbox[t]{5cm}{We collect infinite photons, so the photon noise requirement is meaningless.}\\
n_i^2\lambda^2=\infty & \parbox[t]{5cm}{The inner working angle is infinite.}\\
\end{cases}\end{equation*}
Thus, we conclude that a survey without precursor knowledge will always be limited by photon noise except in the trivial cases noted.  In other words, photon noise considerations impose a larger minimum diameter requirement than the $iwa$. This may seem to be a counter-intuitive result at first: one can posit an ``impossible'' scenario where a small telescope diameter $d$ with a large $iwa$ can still collect the necessary number of photons given sufficient time to image an exo-Earth, but will traditionally be considered $iwa$-limited and unable to image close-in planets that are always inside the $iwa$. However, our photon-noise model treatment accounts for the observing time lost when the exo-Earth is inside the $iwa$, which drives up the observation time required and consequently the minimum telescope diameter. In this sense, our photon noise model already includes the impact of the $iwa$ constraint, leading to this trivial equality. This also means that the previous result for $T$ diverging in section \ref{sec:iwalim:noprior} is irrelevant, because the situation in question never occurs -- the photon noise model will always require a larger telescope diameter -- and this situation of a diverging survey duration is in sense the ``impossible'' posited scenario.

\subsection{Precursor Knowledge}\label{sec:limdetect:prior}
In the case where we have precursor knowledge, when we are limited by photon noise, the necessary minimum diameter is given by:
\begin{equation}d_{\text{noise}}=4SNR_0\left(\frac{3}{4\pi\rho_*\eta_\oplus}\right)^{1/3}\sqrt{\frac{3rN_\oplus^{5/3}}{5RKT\varepsilon}}\tag{\ref{eq:dphotonnoise}}\end{equation}
Note that we use the simple approximation for the summation over $k$.\newline
When we are limited by the inner working angle constraint,
\begin{equation}d_{\text{iwa}}=\frac{n_i D_{\lim}\lambda}{a}\tag{\ref{eq:dminiwa}}\end{equation}
At an intersection between the photon noise and $iwa$ regimes, both $d$-values must be equal.\newline
Let us consider when the situation is limited by the inner working angle constraint. Then,
\begin{equation*}d_\text{iwa}>d_\text{noise}\end{equation*}
\begin{equation}\frac{n_i D_{\lim}\lambda}{a}>4SNR_0\left(\frac{3}{4\pi\rho_*\eta_\oplus}\right)^{1/3}\sqrt{\frac{3rN_\oplus^{5/3}}{5RKT\varepsilon}}.\end{equation}
We rearrange the inequality, for $N_\oplus>0$:
\begin{equation*}\frac{n_i \lambda}{a}\sqrt{\frac{5RKT\varepsilon}{3rN_\oplus^{5/3}}}D_{\lim}>4SNR_0\left(\frac{3}{4\pi\rho_*\eta_\oplus}\right)^{1/3}\end{equation*}
Because cubing preserves order, we cube both sides:
\begin{equation*}\left(\frac{n_i \lambda}{a}\sqrt{\frac{5RKT\varepsilon}{3rN_\oplus^{5/3}}}\right)^{3}D_{\lim}^3>\frac{48SNR_0^3}{\pi\rho_*\eta_\oplus}\end{equation*}
We can expand $D_{\lim}:$
\begin{equation*}\left(\frac{n_i \lambda}{a}\sqrt{\frac{5RKT\varepsilon}{3rN_\oplus^{5/3}}}\right)^{3}\frac{3N_\oplus}{4\pi\rho_*\eta_\oplus}>\frac{48SNR_0^3}{\pi\rho_*\eta_\oplus}\end{equation*}
Because $\eta_\oplus,$ $\rho_*,$ and $\pi$ are all positive, we can multiply by $\frac43\pi\rho_*\eta_\oplus$:
\begin{equation*}\left(\frac{n_i\lambda}{a}\sqrt{\frac{5RKT\varepsilon}{3rN_\oplus^{5/3}}}\right)^{3}N_\oplus>64SNR_0^3\end{equation*}
A cube root also preserves order, so we can take the cube root and simplify:
\begin{equation*}\frac{n_i\lambda}{a}\sqrt{\frac{5RKT\varepsilon}{3rN_\oplus^{5/3}}}N_\oplus^{1/3}>4SNR_0\end{equation*}
\begin{equation*}\frac{n_i\lambda}{a}\sqrt{\frac{5RKT\varepsilon}{3rN_\oplus}}>4SNR_0\end{equation*}
Squaring preserves order over all non-negative numbers, and both sides of this inequality are defined to be positive, so we can square both sides and rearrange:
\begin{equation*}\left(\frac{n_i\lambda}{a}\right)^2\frac{5RKT\varepsilon}{3N_\oplus}>16rSNR_0^2\end{equation*}
\begin{equation*}\frac{n_i^2\lambda^2}{a^2}\frac{5RKT\varepsilon}{3N_\oplus}>16rSNR_0^2\end{equation*}
\begin{equation}\label{eq:iwaLimitedPrior}\frac{5n_i^2\lambda^2RKT\varepsilon}{48rSNR_0^2a^2}>N_\oplus\end{equation}
So, we are limited by the inner working angle for all $N_\oplus$ such that
\begin{equation*}N_\oplus<\frac{5n_i^2\lambda^2RKT\varepsilon}{48rSNR_0^2a^2}\end{equation*}
Similarly, we are limited by photon noise for all $N_\oplus$ such that
\begin{equation*}N_\oplus>\frac{5n_i^2\lambda^2RKT\varepsilon}{48rSNR_0^2a^2}\end{equation*}
and the intersection between the inner working angle and photon noise limited regimes occurs when
\begin{equation}\label{eq:iwaphotonint_noplus}N_\oplus=\frac{5n_i^2\lambda^2RKT\varepsilon}{48rSNR_0^2a^2}\end{equation}

\section{Results}\label{sec:results}

\subsection{Key Equations Summary}

Herein we summarize our key results in deriving the dependence of the minimum telescope diameter on the mission parameter variables under study, given our assumptions. For a direct imaging survey without precursor knowledge, the minimum telescope diameter able to achieve the required yield is \begin{equation}d = \frac{n}{\sqrt{2}}\sqrt[3]{\frac{3N_\oplus}{4\pi\rho_*\eta_\oplus}}\sqrt{1+\sqrt{1+\frac{9216 r^2SNR_0^4 N_{\oplus}^{2}}{25 R^2K^2\varepsilon^2 n^4 T^2\eta_\oplus^2}}}\tag{\ref{eq:dphotonnoprioriwaeta}}\end{equation}
Where $n\equiv n_i\lambda/a.$

For a direct imaging survey with perfect precursor knowledge, the minimum telescope diameter necessary to satisfy the photon noise requirement for a given yield is
\begin{equation}
d =4SNR_0\left(\frac{3}{4\pi\rho_*\eta_\oplus}\right)^{1/3}\sqrt{\frac{3rN_\oplus^{5/3}}{5RKT\varepsilon}}\tag{\ref{eq:dphotonnoise}}
\end{equation}
The minimum telescope diameter necessary to satisfy the inner working angle requirement is
\begin{equation}
d = \frac{n_i\lambda}{a}\sqrt[3]{\frac{3N_\oplus}{4\pi\rho_*\eta_\oplus}}\tag{\ref{eq:dminiwaeta}}
\end{equation}
Solving for the dependence on other variables besides telescope diameter are derived above and not repeated here.

\subsection{When are we limited by iwa angle?}

For no precursor knowledge, the minimum telescope diameter required is always driven by the photon noise requirement, when accounting for time lost due observations taken when the planet is inside the $iwa$, except in trivial situations. In the case of perfect precursor knowledge, the minimum telescope diameter required is limited by the inner working angle requirement when \begin{equation}\frac{5n_i^2\lambda^2RKT\varepsilon}{48rSNR_0^2a^2}>N_\oplus\tag{\ref{eq:iwaLimitedPrior}}\end{equation}

Conversely, the minimum telescope diameter is limited by the photon noise requirement when \begin{equation*}\frac{5n_i^2\lambda^2RKT\varepsilon}{48rSNR_0^2a^2}<N_\oplus\end{equation*}

\section{Discussion}\label{sec:discuss}

We have derived simplified analytic expressions and scaling relations for the telescope diameter as a function of key direct imaging mission parameters such as the occurrence rate of Exo-Earths, the mission yield, survey duration, and other mission properties. We simplified our analytic treatment by assuming identical Sun-like stars with 1 au exo-Earths, with a simplified imaging noise model and other simplifying assumptions.  We now turn to compare our analytic model to more-detailed computational simulations of mission yield calculations and dependencies performed for the HabEx and LUVOIR-B mission concept studies in ${\S}$\ref{sec:HabExLUVOIR}.  Both mission concepts were studied by NASA as input to the Decadal Survey on Astronomy and Astrophysics 2020 \citep{NAP26141}, which recommended the development of a science and technology maturation program leading to the Habitable Worlds Observatory future direct imaging mission of comparable scale to these two mission concepts.  These detailed HabEx and LUVOIR simulations included more complex noise models, target lists and other treatments, and herein we aim to see if these treatments are consistent with our simplified analytic model, and vice-versa.  We then compare our analytic scaling dependencies to the analytical treatment in \citet{2007MNRAS.374.1271A}, and those derived from more detailed computational simulations in \citet{Stark_2014}, in ${\S}$\ref{sec:Agol} and ${\S}$\ref{sec:Stark} respectively.  Finally, we discuss how key mission parameter choices impact Exo-Earth yield, which will be useful in future mission design trade studies to supplement more detailed computational simulations in ${\S}$\ref{sec:trades}.

\subsection{Applications to HabEx \& LUVOIR-B}\label{sec:HabExLUVOIR}

In Table \ref{table:gen}, we list assumed and calculated model parameter values common to both our HabEx and LUVOIR models, and values specific to either HabEx or LUVOIR, such as the telescope diameter, in Tables \ref{table:habex} and \ref{table:luvoir} respectively.  Most values are adopted directly from assumed values either the HabEx and/or LUVOIR reports \citep{theluvoirteam2019luvoir,gaudi2020habitable}, and the Standards and Definitions Team report \citep{Standards}.  For our simplified SNR noise model, we calculate $r'$ through a fixed-origin linear regression fit between $C_{p0}$ and $C_b$ on the EXOSIMS data found in Table 7 of \citet{Standards} for the 9 target stars simulated. This resulted in a value of $r' = 8.75\pm2.29.$ with an $r^2=0.64$ goodness of fit statistic. With the exception of HIP 17651, our model predicts the EXOSIMS exposure time estimates for the remaining 8 targets to $98\pm19\%$.  In other words, our simple SNR photon and background noise model for estimating target exposures times is a reasonable approximation to within $\sim$20\% of the more detailed calculations carried out in EXOSIMS.   By comparison, the Altruistic Yield Optimization tool (AYO) estimates exposure times with an average fractional difference of 56\% compared to the estimated EXOSIMS exposure times \citep[Table 7,][]{Standards}. Given that these two more detailed computational simulations predict exposure times that differ on average by 56\%, and our toy noise model that predicts the EXOSIMS times to within 20\%, our model is thus a reasonable an adequate approximation for our purposes.

Next, $R$ was calculated from Planck's law for each survey:
\begin{align}\label{eq:R}
    R(\nu) &= B_\nu(\nu, T_*)\cdot\frac{1}{h\nu}\cdot4\pi r_*^2\cdot4\pi\cdot \nu \cdot \frac{\Delta \lambda}{\lambda}\\
    &=\frac{32\pi^2r_*^2c\Delta\lambda}{\lambda^4\left(e^{\frac{hc}{\lambda k T_*}} - 1\right)}
\end{align}
Finally, we assume a stellar density of 0.05 Sun-like stars per cubic parsec, which is approximately the stellar mass density of main sequence stars in the Solar Neighborhood excluding mid and late M dwarfs, given that we assume solely Sun-like stars in this work \citep{Chabrier_2001,mamajek_2019}.

\begin{center}
\begin{deluxetable}{cccc}\label{table:gen}
\tablecaption{General values}
\tablehead{\colhead{Variable} & \colhead{Value} & \colhead{Units} & \colhead{Provenance}}
\startdata
    $\rho_*$ & 0.05 & pc$^{-3}$ & CH, MJ\\
    $a$ & $1$ & au & IAU\\
    $\eta_\oplus$ & $0.24$ & \nodata & SDET\\
    $\lambda$ & $500$ & nm & this work\\
    $r'$ & 8.75 & \nodata & this work\\
    $r$ & 9.75 & \nodata & this work\\
    $r_*$ & $6.957\cdot10^{8}$ & m & IAU\\
    $t_*$ & $5772$ & K & IAU\\
    $\varepsilon$ & 0.5 & \nodata & G20\\
    $\lambda / \Delta \lambda$ & 140 & \nodata & G20, L19\\
    $R$ & $1.81218\cdot10^{43}$ & s$^{-1}$ & Eqn \ref{eq:R}\\
    $K$ & $1\cdot10^{-10}$ & \nodata & G20, L19\\\enddata
\tablecomments{$r_*$ and $t_*$ are the radius and temperature of a solar analogue.}
\tablerefs{SDET: \citet{Standards}, CH: \citet{Chabrier_2001}, IAU: \citet{IAU_Res}, MJ: \citet{mamajek_2019}, G20: \citet{gaudi2020habitable}, L19: \citet{theluvoirteam2019luvoir}}
\end{deluxetable}
\end{center}

\begin{center}
\begin{deluxetable}{cccc}\label{table:habex}
\tablecaption{HabEx-specific values}
\tablehead{\colhead{Variable} & \colhead{Value} & \colhead{Units} & \colhead{Provenance}}
\startdata
    $SNR_0$ & 7 & \nodata  & G20 \\
    $N_\oplus$ & 8 & \nodata  & G20 \\
    $d$ & 4 & m  & G20\\
    $T$ & $0.55\cdot2$ & yr  & G20 \\
    $n_i$ & $3$ & \nodata & G20 \\
\enddata
\tablecomments{$T$ was calculated by multiplying the survey duration (two years) by a percent time efficiency ($55\%$), resulting in an estimate of the total on-sky time. $R$ was calculated as described above.}
\tablerefs{G20: \citet{gaudi2020habitable}}
\end{deluxetable}
\end{center}

\begin{figure*}
\includegraphics[scale=0.5]{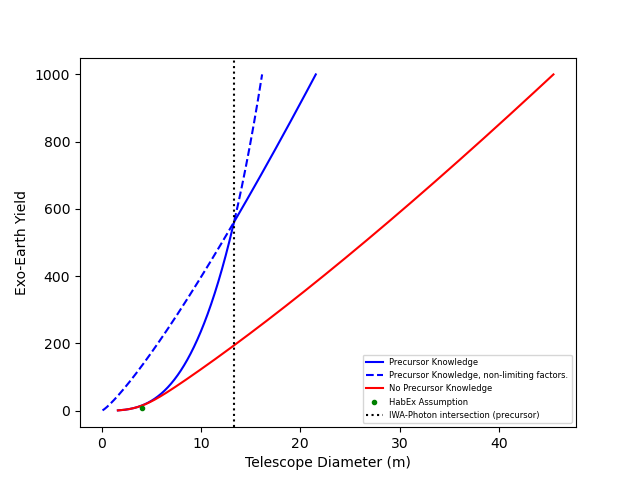}
\includegraphics[scale=0.5]{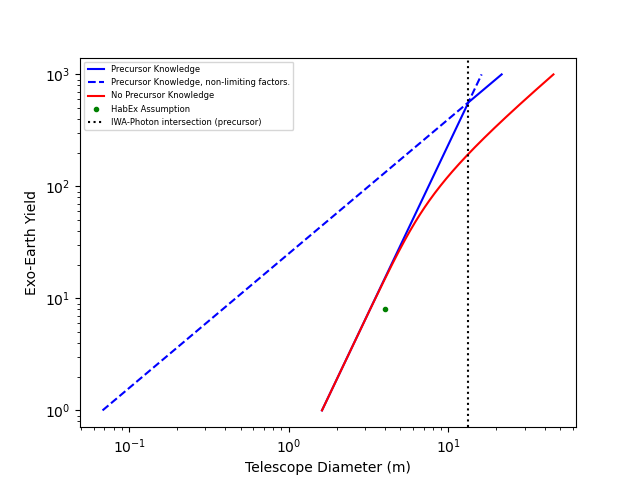}\\
\includegraphics[scale=0.5]{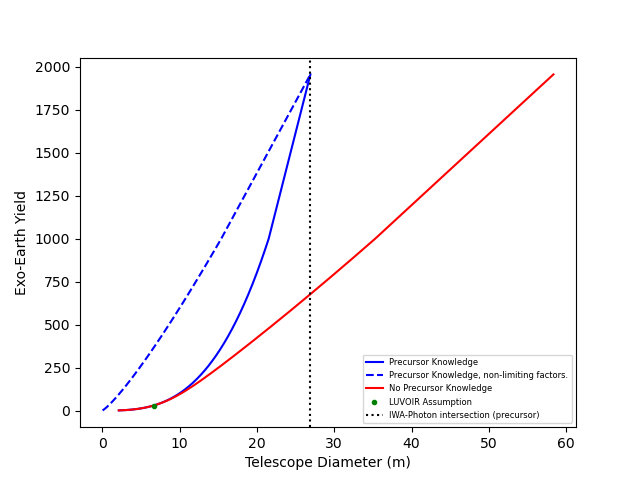}
\includegraphics[scale=0.5]{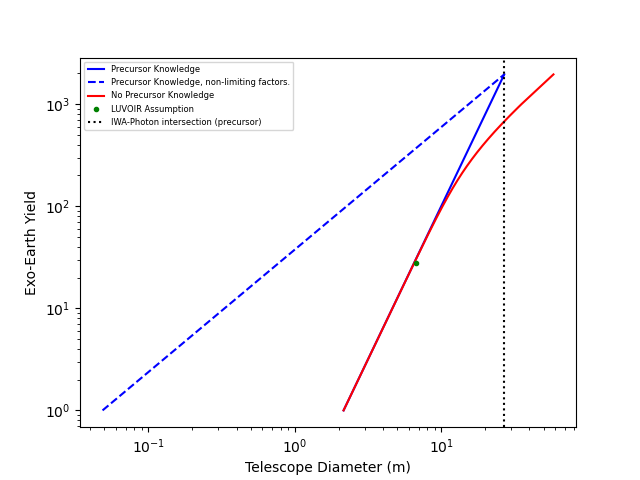}
\caption{The number of Exo-Earth candidates detected as a function of telescope diameter with and without precursor knowledge, comparable to the upper left of Fig. 8 (linear) and 11 (logarithmic) in \citet{Stark_2014}, shown for HabEx with linear (top-left) and logarithmic axes (top-right), and for LUVOIR with linear (bottom-left) and logarithmic (bottom-right) axes.  While the two model curves may look identical, there are differing assumptions for HabEx and LUVOIR mission parameters as detailed in Tables \ref{table:habex} and \ref{table:luvoir}.}
\label{fig:noplus_d}
\end{figure*}

In Figure \ref{fig:noplus_d} we plot the expected Exo-Earth yield as a function of telescope diameter for HabEx and LUVOIR compared to our model.  First, we find that our LUVOIR yield model is in agreement with the more detailed computational simulation yield in \citet{theluvoirteam2019luvoir}, whereas the estimated yield for HabEx is $\sim$50\% lower than our model predicts \citet{gaudi2020habitable}.  Specifically, the HabEx yield estimate of 8 Exo-Earths is indicated with a green dot compares to our estimates of 16 and 15 for the precursor and no precursor knowledge cases at the same telescope diameter. For the LUVOIR yield estimate of 28 Exo-Earths, our model predicts yields of 30 and 30 for the precursor and no precursor knowledge cases at the same telescope diameter, a difference of $<$10\%.  Second, we find that there is no benefit in the Exo-Earth yield from precursor knowledge at this telescope diameter range; of course this ignores additional benefits such as providing contemporaneous mass measurements or orbit determination, and the benefit in survey efficiency which we discuss below.  The benefit of precursor knowledge for Exo-Earth yield is limited to diameters $\gtrsim$10-m for these assumed mission parameter values.

\begin{figure}
\includegraphics[scale=0.5]{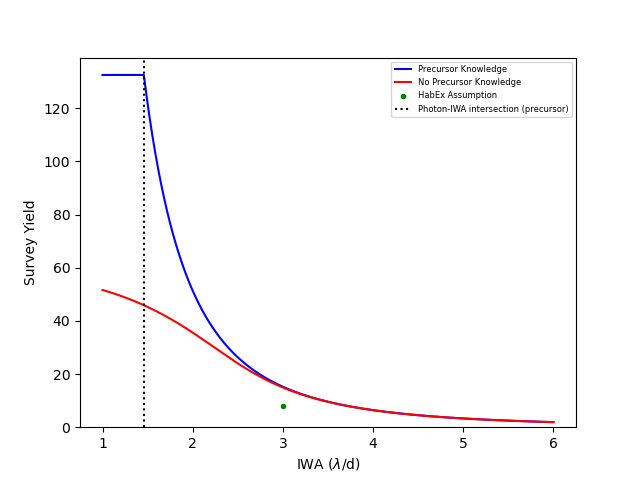}
\includegraphics[scale=0.5]{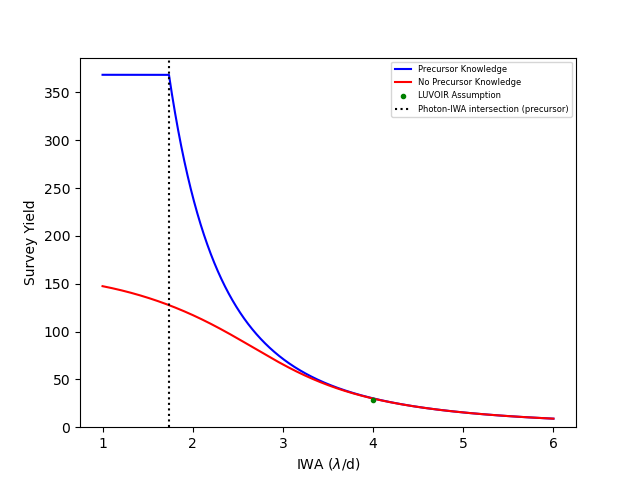}
\caption{The number of ExoEarth Candiates detected vs telescope $iwa$ with and without precursor knowledge, comparable to the upper right of Fig. 8 in \citet{Stark_2014}, for HabEx (top) and LUVOIR (bottom).}
\label{fig:noplus_iwafac}
\end{figure}

In Figure \ref{fig:noplus_iwafac}, we plot the expected Exo-Earth yield as a function of inner-working angle for HabEx and LUVOIR compared to our model. Here we see that there is no benefit from precursor knowledge as a function of inner working angle for the mission parameters and assumed $iwa$.  However, if smaller $iwa$ becomes technically feasible in the future -- e.g. an $iwa=2$ -- then there can be a significant (factor of $\sim$2) yield boost from precursor knowledge.

\begin{figure*}
\includegraphics[scale=0.5]{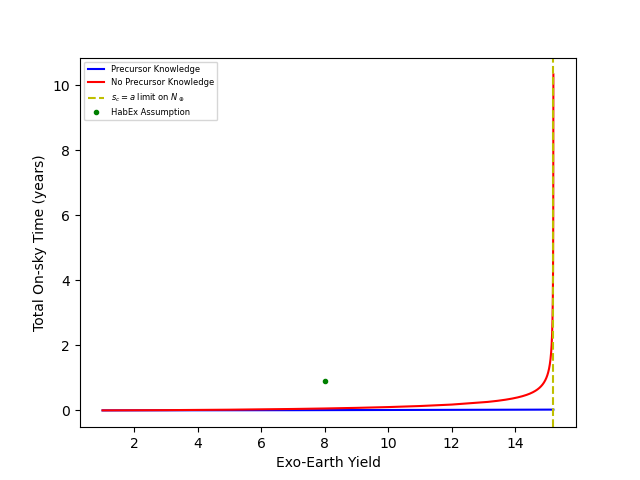}
\includegraphics[scale=0.5]{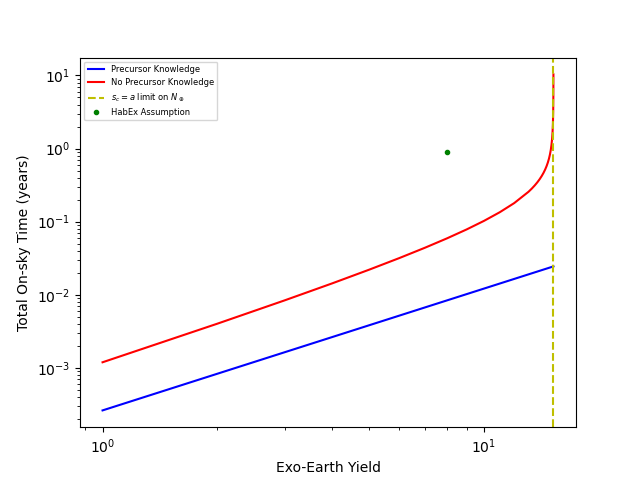}\\
\includegraphics[scale=0.5]{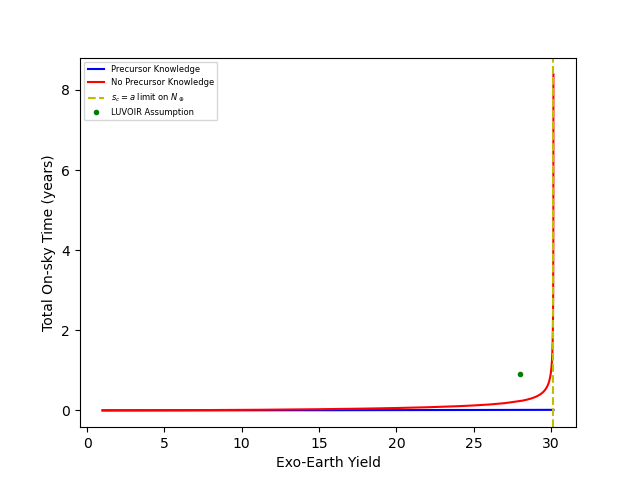}
\includegraphics[scale=0.5]{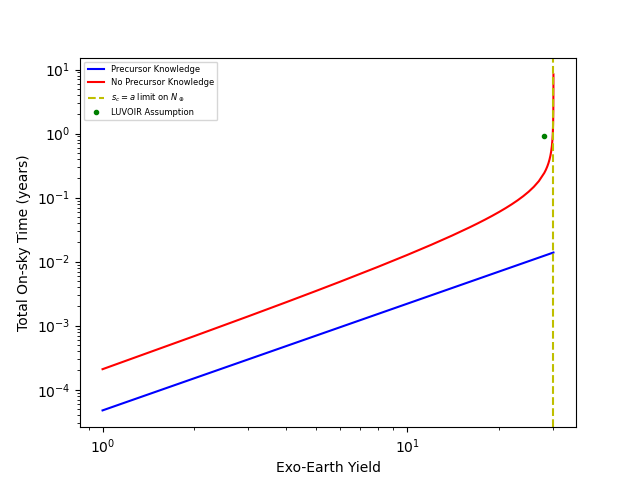}
\caption{The necessary total on-sky time as a function of required survey yield, with and without precursor knowledge. Note that we gain significant time-efficiency improvements from precursor knowledge as we approach the IWA-mandated yield limit. The top panels are for HabEx, and the bottom for LUVOIR. The left panels are for a linear vertical axis, and logarithmic for the right panels. This figure is a flipped-axis version of Fig. 10 in \citet{Stark_2014}.  Note, our survey durations do not include multiple revisits, slew and overhead times, explaining our significantly shorter survey durations.}
\label{fig:noplus_T}
\end{figure*}

Next, in Figure \ref{fig:noplus_T}, we plot the Exo-Earth yield as a function of survey duration for HabEx and LUVOIR compared to our model.  First, we see for the model, the (perfect) precursor knowledge provides a substantial factor of a several reduction in survey time needed to reach the same Exo-Earth yield.  At a yield of 10 Exo-Earths, for HabEx this corresponds to a reduction in on-sky time from 37.37 to 4.42 days (a factor of $\sim$8.5), and for LUVOIR a reduction from 4.65 to 0.8 days (a factor of $\sim$5.7). Note our simulations assume a single visit per target, which can be scaled for multiple visits. For both HabEx and LUVOIR and their expected yields, we see the estimated survey duration is longer than our estimated survey time for a single visit with no precursor knowledge (three times longer in the case of LUVOIR).  This is consistent with needing to account for multiple visits per star. Note in all cases, we also do not account for slews and target acquisition times in calculating on-sky survey duration, which is captured in more detailed computational simulations.

\begin{center}
\begin{deluxetable}{cccc}
\tablecaption{LUVOIR-specific values}\label{table:luvoir}
\tablehead{\colhead{Variable} & \colhead{Value} & \colhead{Units} & \colhead{Provenance}}
\startdata
    $SNR_0$ & 5 & \nodata &  L19\\
    $N_\oplus$ & 28 & \nodata & L19\\
    $d$ & 6.7 & m & L19\\
    $T$ & $0.55\cdot2$ & yr & L19\\
    $n_i$ & $4$ & \nodata & L19 \\
    \enddata
\tablecomments{As before, $T$ was calculated by multiplying the survey duration (two years) by a percent time efficiency ($55\%$). $R$ was calculated as described above.}
\tablerefs{L19: \citet{theluvoirteam2019luvoir}}
\end{deluxetable}
\end{center}

\subsection{Comparison to \citet{2007MNRAS.374.1271A}}\label{sec:Agol}
The assumptions presented in \citet{2007MNRAS.374.1271A} are most closely analogous to those of the photon-noise-limited case without precursor knowledge. \citet{2007MNRAS.374.1271A} finds that $N_\oplus\propto T^{1/3}$ when limited by PSF noise. This paper finds a higher proportionality when the inner working angle requirement is not considered: $N_\oplus\propto T^{3/5}$. The proportionality is less clear when we include the inner working angle requirement: by reparameterizing equation \ref{eq:Nstarpolynomial} with $x=N_*^{2/3}$, we can see that it is a quintic. There is no general formula for the roots of a quintic. We can see from equation \ref{eq:nstarnoprioriwasolve} that the result will be weaker than $T^{3/5}$ for $\beta>0$ ($\beta=0$ represents a lack of the inner working angle requirement, and gives the $3/5$ power), but it won't be a simple power law.

\citet{2007MNRAS.374.1271A} also finds that $N_\oplus\propto SNR_0^{-1}.$ We find a similar exponent when the inner working angle requirement is not considered: $N_\oplus\propto SNR_0^{-6/5}$. The proportionality is less clear when we include the inner working angle requirement, but it will again be weaker than $SNR_0^{-6/5}$ for $\beta>0$, as can be seen from equation \ref{eq:nstarnoprioriwasolve}.

We attribute some of the differences between our results and those presented in \citet{2007MNRAS.374.1271A} to the differences in the assumptions that were made. Specifically, assumptions about stellar spectral type, observation wavelength, semi-major axis, and noise sources differed between our models.  Regarding the first assumption, \citet{2007MNRAS.374.1271A} took into account variable spectral types using the local interstellar mass function, whereas we assumed Solar type stars; we defer to a future work investigating for our models the impact on Exo-Earth yield with stellar mass / spectral type.

\subsection{Comparison to \citet{Stark_2014}}\label{sec:Stark}
As mentioned in the captions for Figures \ref{fig:noplus_d}, \ref{fig:noplus_iwafac}, and \ref{fig:noplus_T}, there are analogous figures in \citet{Stark_2014} and power-law fits to more detailed computational simulations.  For the dependence of Exo-Earth yield on telescope diameter, \citet{Stark_2014} finds a dependence of $N_\oplus \propto d^{1.8}$.  For our model, Equation \ref{eq:dphotonnoprioriwaeta} applies in the photon-noise limited regime with no precursor knowledge:
\begin{equation}d = \frac{n}{\sqrt{2}}\sqrt[3]{\frac{3N_\oplus}{4\pi\rho_*\eta_\oplus}}\sqrt{1+\sqrt{1+\frac{9216 r^2SNR_0^4 N_{\oplus}^{2}}{25 R^2K^2\varepsilon^2 n^4 T^2\eta_\oplus^2}}}\tag{\ref{eq:dphotonnoprioriwaeta}}\end{equation}
This is not a simple power-law that can be inverted for $N_\oplus$, but we can take two simple limits to establish some bounding cases.  First, in the limit that $\frac{9216 r^2SNR_0^4 N_{\oplus}^{2}}{25 R^2K^2\varepsilon^2 n^4 T^2\eta_\oplus^2} \gg 1$, we have that:
\begin{equation}\tag{\ref{eq:noplus_mostbasic}} N_\oplus=\eta_\oplus\sqrt[5]{\frac{125d^6\pi^2\rho_*^2R^3K^3T^3\varepsilon^3}{62208SNR_0^6r^3}}\end{equation}
and we find a shallower dependence on telescope diameter $N_\oplus \propto  d^{\frac{6}{5}}$ than in \citet{Stark_2014}. However, \citet{Stark_2014} does take into account multiple visits, which is in some sense taking into account the impact of $iwa$ and precursor knowledge.  In our model, a second bounding case can be established by the minimum telescope diameter in the $iwa$ limited regime, which is equivalent to Equation \ref{eq:dphotonnoprioriwaeta} in the limit of $\frac{9216 r^2SNR_0^4 N_{\oplus}^{2}}{25 R^2K^2\varepsilon^2 n^4 T^2\eta_\oplus^2} \ll 1$:
\begin{equation}
d = \frac{n_i\lambda}{a}\sqrt[3]{\frac{3N_\oplus}{4\pi\rho_*\eta_\oplus}}\tag{\ref{eq:dminiwaeta}}
\end{equation}
which solving for $N_\oplus$ yields Equation \ref{eq:noplusmaxiwa} with an equality rather than the limit:
\begin{equation}\label{eq:noplusmaxiwa2}N_\oplus = \frac{4\pi\rho_*\eta_\oplus a^3d^3}{3n_i^3\lambda^3}\end{equation}
or a dependence on telescope diameter of $N_\oplus \propto d^3$.  Thus the \citet{Stark_2014} power-law fit lies between these two bounding cases established by our model.

Next, Figure 8 in \citet{Stark_2014} also evaluates the Exo-Earth yield as a function of $iwa$ and finds a dependence of $N_\oplus = 100.95 -78.44\times{iwa}^{0.13}$.  We find a qualitatively similar curve in Figure \ref{fig:noplus_iwafac}, but a different functional form.  From Equation \ref{eq:dphotonnoprioriwaeta}, we assumed $iwa = n_i\lambda / d$, which is encapsulated in our variable $n \equiv n_i\lambda / a$.  To re-express \ref{eq:dphotonnoprioriwaeta} in terms of $iwa$, $iwa = n\:a/d$, and canceling a factor of $d$ from both sides, we have:

\begin{equation}1 = \frac{iwa}{a\sqrt{2}}\sqrt[3]{\frac{3N_\oplus}{4\pi\rho_*\eta_\oplus}}\sqrt{1+\sqrt{1+\frac{9216 r^2SNR_0^4 N_{\oplus}^{2} a^4}{25 R^2K^2\varepsilon^2 iwa^4 d^4 T^2\eta_\oplus^2}}}\label{eq:dphotonnoprioriwaeta_iwa}\end{equation}
This is a non-trivial equation for $N_\oplus(iwa)$, but we can consider two limiting case power laws as we did previously.  First, in the limit that $\frac{9216 r^2SNR_0^4 N_{\oplus}^{2} a^4}{25 R^2K^2\varepsilon^2 iwa^4 d^4 T^2\eta_\oplus^2} \gg 1$, we have that $N_\oplus$ is independent of $iwa$, which corresponds to the photon-noise limited regime as one would expect:

\begin{equation}
N_\oplus = \eta_\oplus \sqrt[5]{\frac{125\pi^2 \rho_*^3 R^3K^3\epsilon^3T^3d^6}{62208 r^3 SNR_0^6}}
\end{equation}

Second, in the limit $\frac{9216 r^2SNR_0^4 N_{\oplus}^{2} a^4}{25 R^2K^2\varepsilon^2 iwa^4 d^4 T^2\eta_\oplus^2} \ll 1$, we have in the inner working angle limited regime with no precursor knowledge: 
\begin{equation}
N_\oplus = \frac{\eta_\oplus 4\pi\rho_* a^3}{3 iwa^3}
\end{equation}
This is quite a large range from our two limiting scenarios, and thus depending on the mission parameter choices, the Exo-Earth yield can range from very little to a very steep dependence on the mission $iwa$.  To support this conclusion, we note Figure 8 in \citet{Stark_2015} (not to be confused with the similar Figure 8 in \citet{Stark_2014}) evaluates a dependence of $iwa^{-0.98}$ for that assumed mission architecture, whereas the dependence on $iwa$ in \citet{gaudi2020habitable} is relatively flat.

Finally, in Figure 10, \citet{Stark_2014} investigates the survey Exo-Earth yield as a function of total on-sky time, the reciprocal of our Figure \ref{fig:noplus_T}, and finds that the yield scales as mission duration $T^{0.41}$.  Again, from Equation \ref{eq:dphotonnoprioriwaeta}, we can solve for $T$ as a function of $N_\oplus$, but not in the inverse.  For the latter we must use the same two prior approximations:

\begin{equation}
T = \frac{48\times3^{\frac{2}{3}} SNR^2 N_\oplus^{\frac{5}{3}}}{ 5RK\epsilon d (4\pi\rho_*)^{\frac{1}{3}} \eta_\oplus^{\frac{4}{3}} \sqrt{ d^2 (4\pi \rho_* \eta_\oplus)^{\frac{2}{3}} - n^2 (3 N_\oplus)^{\frac{2}{3}}  }}
\end{equation}
In the limit that $\frac{9216 r^2SNR_0^4 N_{\oplus}^{2}}{25 R^2K^2\varepsilon^2 n^4 T^2\eta_\oplus^2} \gg 1$, which can be thought of the short survey duration limited case, we again derive:
\begin{equation}\tag{\ref{eq:noplus_mostbasic}} N_\oplus=\eta_\oplus\sqrt[5]{\frac{125d^6\pi^2\rho_*^2R^3K^3T^3\varepsilon^3}{62208SNR_0^6r^3}}\end{equation}
and in the limit of $\frac{9216 r^2SNR_0^4 N_{\oplus}^{2}}{25 R^2K^2\varepsilon^2 n^4 T^2\eta_\oplus^2} \ll 1$, which can be thought of as the long survey duration case, we again derive:
\begin{equation}\tag{\ref{eq:noplusmaxiwa2}}N_\oplus = \frac{4\pi\rho_*\eta_\oplus a^3d^3}{3n_i^3\lambda^3}\end{equation}
In other words, for long-enough survey durations, you run out of targets to image and the survey yield asymptotes to be independent of survey duration; this is the vertical asymptote in Figure \ref{fig:noplus_T}.  Specifically then, we find that $N_\oplus \propto T^{3/5}$ in the short survey duration regime, slightly steeper than the power law in \citet{Stark_2014},although without the zero point offset.

\subsection{Evaluating the Dependence of Exo-Earth Yield on Mission Parameter Choices, Trades, and Precursor Knowledge}\label{sec:trades}

In the previous section, we evaluated the dependence of Exo-Earth yield on the telescope diameter $d$, the $iwa$ and on sky survey duration $T$ in the absence of precursor knowledge, which is given by Equation  \ref{eq:dphotonnoprioriwaeta} in a simplifying limit when we are inner working angle limited and goes as Equation \ref{eq:noplus_mostbasic}:
\begin{equation}d = \frac{n}{\sqrt{2}}\sqrt[3]{\frac{3N_\oplus}{4\pi\rho_*\eta_\oplus}}\sqrt{1+\sqrt{1+\frac{9216 r^2SNR_0^4 N_{\oplus}^{2}}{25 R^2K^2\varepsilon^2 n^4 T^2\eta_\oplus^2}}}\tag{\ref{eq:dphotonnoprioriwaeta}}\end{equation}
\begin{equation}\tag{\ref{eq:noplus_mostbasic}} N_\oplus=\eta_\oplus\sqrt[5]{\frac{125d^6\pi^2\rho_*^2R^3K^3T^3\varepsilon^3}{62208SNR_0^6r^3}}\end{equation}

In the case of precursor knowledge, there is a gain in yield derived in both the scenarios considered in equations \ref{eq:noplusprior} and \ref{eq:noplus_prior_iwa}:
\begin{equation}\tag{\ref{eq:noplusprior}}
    N_\oplus=\sqrt[5]{\frac{125\pi^2 d^6 R^3 K^3 T^3 \varepsilon^3 \rho_*^2 \eta_\oplus^2}{62208 r^3 SNR_0^6}}
\end{equation}
\begin{equation}\tag{\ref{eq:noplus_prior_iwa}}
    N_\oplus = \sqrt[5]{\frac{125\pi^2 d^6 R^3 K^3 T^3\varepsilon^3 \rho_*^2 \eta_\oplus^2}{62208r^3SNR_0^6}}
\end{equation}

All of the factors are identical to the no precursor knowledge case with one exception: the Exo-Earth yield for the precursor knowledge cases have a missing factor of $\eta_\oplus$, which is $<1$ and is thus a gain to $N_\oplus$. This yield gain can be thought of as a direct consequence of surveying only $N_\oplus$ stars instead of $N_\oplus\eta_\oplus = N_*$ stars as is the case with no precursor knowledge, and is thus primarily realized through enabling a shorter survey duration in the inner working angle limited regime represented by Equations \ref{eq:noplus_mostbasic}, \ref{eq:noplusprior}, and \ref{eq:noplus_prior_iwa}.  

While we find that, as intuitively expected, the Exo-Earth yield of a mission with no precursor knowledge scaled with $\eta_\oplus$, as also captured as an approximately linear relation in Figure~14 of \citet{Stark_2015}, this is not the case for a survey with precursor knowledge as seen in Equations \ref{eq:noplus_prior_iwa} and \ref{eq:noplusprior}.  In the case of precursor knowledge, the dependence on $\eta_\oplus$ is a much shallower power law of $\frac{2}{5}$.  In other words, precursor knowledge, such as might be obtained by ground-based precise radial velocities or astrometry, reduces the sensitivity and thus the risk of the yield of a future direct imaging mission to our current knowledge of $\eta_\oplus$ and its uncertainty \citep[e.g., ][and references therein]{2021AJ....161...36B,2019MNRAS.483.4479Z}. 

Next, throughout this work we have assumed ``perfect'' precursor knowledge, when astrometry and radial velocities in general will provide incomplete knowledge of exoplanet systems, either from the lack of a known inclination for radial velocities, or limits to exoplanet mass sensitivity for both techniques that are currently both well above the Earth-mass regime. So the actual impact of precursor knowledge will lie somewhere between our two limiting scenarios of whether or not precursor knowledge is available for a future direct imaging mission.  We have now quantified that benefit analytically as presented herein in terms of Exo-Earth yield and survey efficiency, in support of community evaluations of EPRV precursor surveys in \citet{Crass} and the numerical simulations in \citet{morgan}. Even without ``perfect'' precursor knowledge, more massive planets discovered with precursor radial velocities located in the HZ of target stars can dynamically preclude the presence of HZ Exo-Earths in a given system \citep{Hill2018,Kane2019,Kane2020}, which can in turn help optimize target selection, HZ exo-moons aside \citep{moon1,moon2}.  In addition to mass sensitivity, there is also a need for ephemerides refinement of known more massive planets in the systems of interest discovered with the radial velocity technique, regardless of orbital semi-major axis, to more accurately forecast orbital phase at the imaging epochs to aid in planet identification for what will hopefully be bountiful multi-planet systems when imaged \citep[e.g., ``which is which,''][]{Kane2009}.

Finally, while we have shown how the Exo-Earth yield scales with $iwa$,telescope diameter $d$, survey duration $T$, $\eta_\oplus$, the same can also be done for $SNR_0$, flux contrast $K$, spectral resolution $R$, approximate noise model enhancement factor $r$, planet semi-major axis $a$, target stellar density $\rho_*$, and telescope throughput efficiency $\epsilon$ by differentiating equations presented herein, or through point comparison deltas.  This can potentially be very useful in quick ``rules-of-thumb'' in the coming decade's trade studies in mission and telescope design, and instrument parameters.

\section{Conclusions}
\label{sec:concl}

We have calculated simple analytic expressions for the yield of a future flagship direct imaging mission such as Habitable Worlds Observatory as a function of various key mission design parameters under the assumption of identical and uniformly distributed Sun-like stars. We find that the HabEx and LUVOIR mission concept yield simulations of Earth-like planets are consistent with our analytic model, with little increase in yield from precursor knowledge. However, the benefit from precursor knowledge can increase greatly for larger-yield or larger telescope diameter surveys, or for surveys that require higher SNR and spectral resolution than base-lined for the HabEx and LUVOIR mission concepts. Additionally, we find that precursor knowledge reduces the mission risk (sensitivity) to our Exo-Earth yield given our current knowledge of $\eta_\oplus$ and its uncertainty. Next, we find that the survey efficiency is greatly enhanced by precursor knowledge such as can be provided by extremely precise radial velocities and astrometry, consistent with precursor detailed computational simulations. We also find qualitatively similar agreement to HabEx and LUVOIR yield estimates, and for the dependence of yield of several key mission parameters from more detailed computational simulations. These consistent results provide an analytic check on these more detailed simulations. We have provided a set of relations that allow for fast estimates of the analytic dependence of Exo-Earth yield on key mission, telescope and instrument parameters, both in the bounding cases of no precursor knowledge and full precursor knowledge. In the future, we could explore modifying our analytical model to include a range of spectral types and semi-major axes, as well as for a range of different planet populations.

\section*{Acknowledgements}

PPP and JEB contributed equally to this work. As a high school research intern from 2019-2022, JEB derived most of the analytic relations under the analytic framework and mentorship provided by PPP. JEB developed all of the software for generating figures. PPP performed most of the writing and organization. SE provided initial investigations in a prior year as a summer research intern.  SRK, RM and EP provided detailed feedback and expert guidance to the development of the research.

We thank Karl Stapelfeldt, Eric Mamajek, Scott Gaudi, Thayne Currie, Bryson Cale, and Chris Stark for useful conversations leading to motivating this research and feedback on this analysis over nearly a decade from its initial concept formulation at the start of the HabEx mission concept study. PPP was a member of and acknowledges support from the HabEx STDT and the Standards and Definitions Team.

JEB would like to acknowledge Dr. John E. Berberian, Sr., for the derivation of the better approximation for the finite power-law summation relative to the integral limit.\newline
PPP would like to acknowledge support from NASA (Exoplanet Research Program Award \#80NSSC20K0251, TESS Cycle 3 Guest Investigator Program Award \#80NSSC21K0349, JPL Research and Technology Development, and Keck Observatory Data Analysis) and the NSF (Astronomy and Astrophysics Grants \#1716202 and 2006517), and the Mt Cuba Astronomical Foundation.  



\section{Appendix A}\label{sec:appA}
\subsection{Derivation of $\eta_\oplus$ dependence of $\sum_{k=1}^{N_*} k^{2/3}$.}
Because of the properties of telescoping series, we know that: \begin{align*}(N_*+1)^{p+1} - 1&=\sum_{k=1}^{N_*}(k+1)^{p+1}-k^{p+1}\\&=\sum_{k=1}^{N_*}k^{p+1}\left(\left(1+\frac1k\right)^{p+1} - 1\right)\end{align*}\newline
Using a binomial series expansion, we get:
\begin{align*}
    (N_*+1)^{p+1} - 1&=\sum_{k=1}^{N_*}k^{p+1}\left(-1+\sum_{\ell=0}^{\infty} {{p+1}\choose{\ell}}k^{-\ell}\right)\\
    &=\sum_{k=1}^{N_*}k^{p+1}\sum_{\ell=1}^{\infty} {{p+1}\choose{\ell}}k^{-\ell}\\
    &=\sum_{k=1}^{N_*}\sum_{\ell=1}^{\infty} {{p+1}\choose{\ell}}k^{p+1-\ell}\\
    &=\sum_{\ell=1}^{\infty} {{p+1}\choose{\ell}}\sum_{k=1}^{N_*}k^{p+1-\ell}
\end{align*}\newline
We now define $S_n$ such that \begin{equation*}S_n=\sum_{k=1}^{N_*} k^n.\end{equation*}
Thus, \begin{align*}
    (N_*+1)^{p+1} - 1&=\sum_{\ell=1}^{\infty} {{p+1}\choose{\ell}}S_{p+1-\ell}\\
    (N_*+1)^{p} - 1&=\sum_{\ell=1}^{\infty} {{p}\choose{\ell}}S_{p-\ell}\\
    (N_*+1)^{p-1} - 1&=\sum_{\ell=1}^{\infty} {{p-1}\choose{\ell}}S_{p-1}\\
    &\boldsymbol{\cdot}\\&\boldsymbol{\cdot}\\&\boldsymbol{\cdot}
\end{align*}\newline
A useful representation of this equality is the multiplication of a matrix with a vector.\newline\newline
$\begin{bmatrix}
(N_*+1)^{p+1}-1\\(N_*+1)^{p}-1\\(N_*+1)^{p-1}-1\\\boldsymbol{\cdot}\\\boldsymbol{\cdot}\\\boldsymbol{\cdot}
\end{bmatrix}=
\begin{bmatrix}
{{p+1}\choose{1}}&{{p+1}\choose{2}}&{{p+1}\choose{3}}&\boldsymbol{\cdot}&\boldsymbol{\cdot}&\boldsymbol{\cdot}\\
0&{{p}\choose{1}}&{{p}\choose{2}}\\
0&0&{{p-1}\choose{1}}\\\boldsymbol{\cdot}&&&\boldsymbol{\cdot}\\\boldsymbol{\cdot}&&&&\boldsymbol{\cdot}\\\boldsymbol{\cdot}&&&&&\boldsymbol{\cdot}
\end{bmatrix}
\begin{bmatrix}
S_{p}\\S_{p-1}\\S_{p-2}\\\boldsymbol{\cdot}\\\boldsymbol{\cdot}\\\boldsymbol{\cdot}
\end{bmatrix}$\newline\newline
We now assume that a $2\text{x}2$ matrix is sufficient to approximate this within acceptable error. We will verify this assumption for $p=\frac{2}{3}$ later.\newline\newline
$\begin{bmatrix}
(N_*+1)^{p+1}-1\\(N_*+1)^{p}-1
\end{bmatrix}\approx
\begin{bmatrix}
{{p+1}\choose{1}}&{{p+1}\choose{2}}\\
0&{{p}\choose{1}}\\
\end{bmatrix}
\begin{bmatrix}
S_{p}\\S_{p-1}
\end{bmatrix}$\newline\newline
We now apply Cramer's rule to solve for $S_p$:
\begin{align*}
S_p&=\frac{\begin{vmatrix}
(N_*+1)^{p+1}-1&{{p+1}\choose{2}}\\
(N_*+1)^{p}-1&{{p}\choose{1}}
\end{vmatrix}}{\begin{vmatrix}
{{p+1}\choose{1}}&{{p+1}\choose{2}}\\
0&{{p}\choose{1}}\\
\end{vmatrix}}\\\\
&=\frac{p\left((N_*+1)^{p+1}-1\right)-\frac{p(p+1)}{2}\left((N_*+1)^{p}-1\right)}{p(p+1)}\\
&=\frac{(N_*+1)^{p+1}}{p+1}-\frac{(N_*+1)^{p}}{2}+\left(\frac{1}{2}-\frac{1}{p+1}\right)
\end{align*}\newline
This approximation of $\sum_{k=1}^{N_*}k^p$ for $p=\frac{2}{3}$ has an error of $1.118074\%$ at $N_*=1,$ and an error of $0.000083\%$ at $N_*=1000,$ and as $N_*$ grows, the error continues to decrease. For the purposes of this paper, this error is insignificant.\newline
So, \begin{equation}\label{eq:sumapprox}\sum_{k=1}^{N_*}k^{2/3}\approx\frac{3(N_*+1)^{5/3}}{5}-\frac{(N_*+1)^{2/3}}{2}-\frac{1}{10}\end{equation}

\section{Appendix B}\label{sec:appB}
\subsection{Derivation of Time Fraction Usable}\label{sec:appB:cases}
We assume that the exoplanet's orbit is circular, with some inclination $i$ relative to the viewer. This will make one on-sky axis fore-shortened, by a uniform random factor of $1\geq\cos i\geq0.$ From the viewer's perspective, the exoplanet traces an ellipse, described by the equations \begin{align*}x_p(t)&=a\cos(t)\\y_p(t)&=a\sin(t)\cos i\\\end{align*}
where $a$ is the semi-major axis of the planet. Note that we arbitrarily chose to shorten the $y_p$ dimension; because the viewer's perspective can be rotated.

We now project the inner working angle of the telescope onto that ellipse. This creates a circle of radius \begin{equation}\label{eq:scdef}s_c=D_k\cdot\text{iwa}.\end{equation} A usable observation is one that occurs when $x_p^2+y_p^2>s_c^2$ holds; that is, when the exoplanet in question is outside the circle.
The fraction of time usable will vary depending on the relationship between these variables. To ensure that we have explored all options, we will refer to the following case table.
\begin{table}
    \centering
    \caption{Cases for time fraction usable}\label{tab:casetable}
    \begin{tabular}{|c|c|c|c|c|c|c|c|c|c|}
    \hline
    Axis & \multicolumn{9}{|c|}{Value Relative to $s_c$}\\
    \hline
    $a$ & $<$ & $=$ & $>$ & $<$ & $=$ & $>$ & $<$ & $=$ & $>$ \\
    $a\cos i$ & $<$ & $<$ & $<$ & $=$ & $=$ & $=$ & $>$ & $>$ & $>$ \\
    \hline
    Case & A & B & C & \nodata & D & E & \nodata & \nodata & F \\
    \hline
    \end{tabular}
\end{table}
We do not enumerate the cases that violate the restriction that $\cos i\leq1.$
\subsubsection{Derivation of Time Fraction Usable - Case A}\label{sec:appB:cases:a}
Both axes of the ellipse, $a$ and $a\cos i$, are less than $s_c.$ This means that the ellipse is completely enclosed inside the circle. We can always say that $x_p^2+y_p^2<s_c^2,$ so we cannot make any usable observations. The usable time fraction for this case is $0.$
\subsubsection{Derivation of Time Fraction Usable - Case B}\label{sec:appB:cases:b}
This time, $a=s_c,$ but $a\cos i<s_c.$ The ellipse is tangent to the circle at two points, but it is still completely enclosed by the circle. We can always say that $x_p^2+y_p^2\leq s_c^2,$ so we cannot make any usable observations. The time fraction usable is $0.$
\subsubsection{Derivation of Time Fraction Usable - Case C}\label{sec:appB:cases:c}
This one is the most difficult case. $a>s_c>a\cos i,$ so the expression $x_p^2+y_p^2>s_c^2$ sometimes holds, so we only sometimes get usable observations. Fortunately, we can derive the fraction of time usable for this case.

Substituting our expression for $y_p$ into that inequality:
\begin{equation*}x_p^2+a^2\sin^2\left(t\right)\cos^2 i>s_c^2\end{equation*}
We can use a Pythagorean identity to get everything in terms of $x_p$:
\begin{equation*}x_p^2+\cos^2 i\left(a^2-a^2\cos^2\left(t\right)\right)>s_c^2\end{equation*}
\begin{equation*}x_p^2+\cos^2 i\left(a^2-x_p^2\right)>s_c^2\end{equation*}
Solving for $x_p^2$ in terms of the other variables:
\begin{equation*}x_p^2-x_p^2\cos^2 i>s_c^2-a^2\cos^2 i\end{equation*}
\begin{equation*}x_p^2\left(1-\cos^2 i\right)>s_c^2-a^2\cos^2 i\end{equation*}
\begin{equation*}x_p^2>\frac{s_c^2-a^2\cos^2 i}{1-\cos^2 i}\end{equation*}
To ensure that the last step - dividing by $1-\cos^2 i$ - was valid, we can examine the available values for $\cos^2 i.$ Recall that in this case, $a>s_c>a\cos i.$ If $\cos i=1,$ then $a=a\cos i$, which violates the base assumption for this case. So, $\cos i<1,$ and $1-\cos^2 i>0.$
Moving on, we can take the square root of both sides:
\begin{equation*}x_p>\sqrt{\frac{s_c^2-a^2\cos^2 i}{1-\cos^2 i}}\end{equation*}
\begin{center}
    OR
\end{center}
\begin{equation*}x_p<-\sqrt{\frac{s_c^2-a^2\cos^2 i}{1-\cos^2 i}}\end{equation*}
At the intersections, 
\begin{equation*}x_p=\pm\sqrt{\frac{s_c^2-a^2\cos^2 i}{1-\cos^2 i}}\end{equation*}
These are the x-values for the intersections, but it would be useful to get y-values too.
\begin{align*}
y_p^2&=a^2\cos^2 \left(t\right)\cos^2 i\\
&=\cos^2 i\left(a^2-a^2\sin^2\left(t\right)\right)\\
&=\cos^2 i\left(a^2-x^2\right)
\end{align*}  
Again, taking the square root,
\begin{equation*}y_p=\pm\cos i\sqrt{a^2-\frac{s_c^2-a^2\cos^2 i}{1-\cos^2 i}}\end{equation*}
Notably, both the $x_p$ and $y_p$ have $\pm$ symmetry. An example of this is shown below, with the usable times highlighted in red.

\begin{figure}
    \includegraphics[scale=0.25]{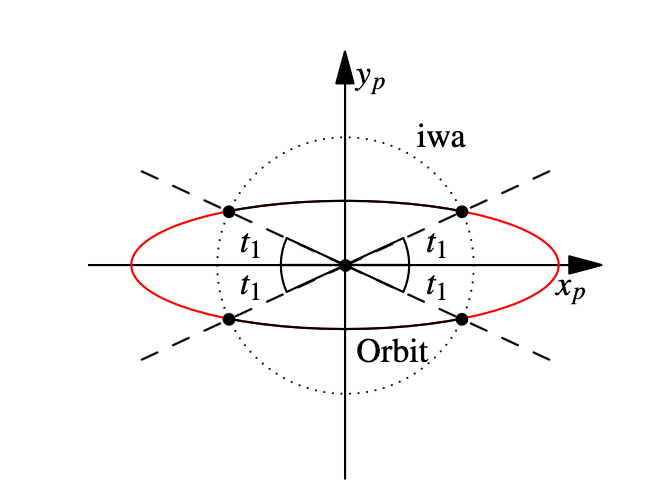}
    \caption{Example of Case C orbit, usable times are highlighted in red.}
\end{figure}

Because of the symmetry, all the angles will be the same. We will call that angle measure $t_1$. We can easily find $t_1$ from the $x_p$ value at one of the intersections. For simplicity's sake, we will choose the positive $x_p$ intersection value.
\begin{equation*}x_p=a\cos\left(t_1\right)=\sqrt{\frac{s_c^2-a^2\cos^2 i}{1-\cos^2 i}}\end{equation*}
\begin{equation*}\cos\left(t_1\right)=\frac{1}{a}\sqrt{\frac{s_c^2-a^2\cos^2 i}{1-\cos^2 i}}\end{equation*}
\begin{equation*}t_1=\cos^{-1}\left(\frac{1}{a}\sqrt{\frac{s_c^2-a^2\cos^2 i}{1-\cos^2 i}}\right)\end{equation*}
Where $\cos^{-1}$ denotes the inverse cosine function. Because there are four such angles, we should multiply this by four.
\begin{equation*}4t_1=4\cos^{-1}\left(\frac{1}{a}\sqrt{\frac{s_c^2-a^2\cos^2 i}{1-\cos^2 i}}\right)\end{equation*}
This gives us the total angle measure in which usable observations can be made. We want this as a fraction of the $2\pi,$ though. We will assume that the angle measure is equivalent to time. Then, the time fraction usable $t_f$ would be:
\begin{align*}t_f&=\frac{4}{2\pi}\cos^{-1}\left(\frac{1}{a}\sqrt{\frac{s_c^2-a^2\cos^2 i}{1-\cos^2 i}}\right)\\&=\frac{2}{\pi}\cos^{-1}\left(\frac{1}{a}\sqrt{\frac{s_c^2-a^2\cos^2 i}{1-\cos^2 i}}\right)\end{align*}
So, the usable fraction of time for this case is \begin{equation}\label{eq:timefrachard}\frac{2}{\pi}\cos^{-1}\left(\frac{1}{a}\sqrt{\frac{s_c^2-a^2\cos^2 i}{1-\cos^2 i}}\right).\end{equation}
\subsubsection{Derivation of Time Fraction Usable - Case D}\label{sec:appB:cases:d}
Because $a=a\cos i=s_c,$ $x_p^2+y_p^2=s_c^2\cos^2\left(t\right)+s_c^2\sin^2\left(t\right),$ which is always $s_c^2.$ Unfortunately, $s_c^2>s_c^2$ is never true, so we will never get usable observations for this case. The fraction of usable time is $0.$
\subsubsection{Derivation of Time Fraction Usable - Case E}\label{sec:appB:cases:e}
$a>s_c=a\cos i,$ so the ellipse is tangent to the circle at two points, and at all other times it is outside the circle. So, we can make usable observations on the exoplanet at all times except two. Because both those points are infinitesimally small, the fraction of time usable for this case is $1.$
\subsubsection{Derivation of Time Fraction Usable - Case F}\label{sec:appB:cases:f}
Both $a$ and $a\cos i$ are greater than $s_c.$ This means that the inner-working-angle circle is completely enclosed inside the orbit's ellipse, so all observations will be usable. The fraction of time usable is $1.$
\subsection{Derivation of Average Time Fraction Usable}\label{sec:appB:avg}
Because we can get no usable observations with $s_c\geq a,$ we require an $\text{iwa}$ such that $s_c<a$ for all targets.\newline\newline
For simplicity, it would be useful to have an average of the usable time fractions. We can achieve this by integrating the fraction of time usable across all values of $\cos i.$ We can do this without weighting because $\cos i$ is a uniform random variable.\newline\newline
While $0\leq\cos i<s_c/a,$ $a\cos i<s_c<a,$ (Case C) so the usable time fraction is \begin{equation*}\frac{2}{\pi}\cos^{-1}\left(\frac{1}{a}\sqrt{\frac{s_c^2-a^2\cos^2 i}{1-\cos^2 i}}\right).\end{equation*}
While $\cos i=s_c/a,$ $a\cos i=s_c<a,$ (Case E) so the usable time fraction is $1.$\newline\newline
While $s_c/a<\cos i\leq1,$ $s_c<a\cos i\leq a,$ (Case F) so the usable time fraction is $1.$\newline\newline
So, our integral will be
\begin{align*}\lim_{\kappa\to\left(\frac{s_c}{a}\right)^{-}}&\int_0^{\kappa} \frac{2}{\pi}\cdot\arccos\left(\frac1a\cdot\sqrt{\frac{s_c^2-a^2 \cos^2i} {1-(\cos^2i)}} \right) d(\cos i)\\+&\int_{s_c/a}^1 1 \,d(\cos i)\end{align*}
Integrating $1$ is trivial, but the arccosine might be harder. So, we turn to Mathematica.
.
Evaluating the Mathematica expression \begin{verbatim}
Limit[Integrate[2 ArcCos[Sqrt[(sc^2 - 
(a^2)(x^2))/(1 - x^2)]/a]/Pi, x], x->0]
\end{verbatim} 
yields \begin{equation*}\frac{2\sqrt{-s_c^2}\sqrt{1-\frac{s_c^2}{a^2}}\ln\left(a\sqrt{-s_c^2}\right)}{\sqrt{s_c^2}\pi}\end{equation*}
This will be simplified in a the "Lower bound" section.\newline
If we evaluate the expression
\begin{verbatim}Limit[Integrate[2 ArcCos[Sqrt[
(sc^2 - (a^2)(x^2))/(1 - x^2)]/a]/Pi, x], 
x -> sc/a, Direction -> "FromBelow"]\end{verbatim} 
we get the result \begin{align*}
\frac{s_c}{a}+\frac{\sqrt{-as_c}}{\pi\sqrt{\frac{a^3s_c}{a^2-s_c^2}}\sqrt{a^2-s_c^2}}&\left(a\ln\left(1-\frac{s_c}{a}\right)-a\ln(s_c(a-s_c))-a\ln\left(\frac{a+s_c}{a}\right)\right.\\
&\left.\;+\;2\sqrt{a^2-s_c^2}\ln(as_c)+a\ln\left(-s_c(a+s_c)\right)\right).\\
\end{align*}
This will be simplified in the "Upper bound section."
\subsubsection{Upper bound}\label{sec:appB:avg:upper}
Direct from Mathematica, with no simplifications:
\begin{align*}
\frac{s_c}{a}+\frac{\sqrt{-as_c}}{\pi\sqrt{\frac{a^3s_c}{a^2-s_c^2}}\sqrt{a^2-s_c^2}}&\left(a\ln\left(1-\frac{s_c}{a}\right)-a\ln(s_c(a-s_c))-a\ln\left(\frac{a+s_c}{a}\right)\right.\\
&\left.\;+\;2\sqrt{a^2-s_c^2}\ln(as_c)+a\ln\left(-s_c(a+s_c)\right)\right).\\
\end{align*}
We know that $a\cos i \geq 0,$ and $s_c>a\cos i,$ so $s_c>0.$ Also, $a>0$ because the planet must orbit at a nonzero distance. Thus, $ab>0$ and $\sqrt{as_c}\neq0,$ so we can cancel it, and pull a factor of $a$ out of the denominator's square root.
\begin{align*}
\frac{s_c}{a}+\frac{\sqrt{-1}}{a\pi\sqrt{\frac{1}{a^2-s_c^2}}\sqrt{a^2-s_c^2}}&\left(a\ln\left(1-\frac{s_c}{a}\right)-a\ln(s_c(a-s_c))-a\ln\left(\frac{a+s_c}{a}\right)\right.\\
&\left.\;+\;2\sqrt{a^2-s_c^2}\ln(as_c)+a\ln\left(-s_c(a+s_c)\right)\right)\\\\ 
\end{align*} 
We know that $a > s_c$, so $a^2-s_c^2\neq0$. So, we can cancel a $\sqrt{a^2-s_c^2}$ in the denominator. Also, $\sqrt{-1}=i.$
\begin{align*}
\frac{s_c}{a}+\frac{i}{a\pi}&\left(a\ln\left(1-\frac{s_c}{a}\right)-a\ln(s_c(a-s_c))-a\ln\left(\frac{a+s_c}{a}\right)\right.\\
&\left.\;+\;2\sqrt{a^2-s_c^2}\ln(as_c)+a\ln\left(-s_c(a+s_c)\right)\right)\\\\
\end{align*} 
Because the exoplanet must orbit its star at some nonzero distance, we know that $a\neq0$. So, we can cancel a factor of $a.$
\begin{align*}
\frac{s_c}{a}+\frac{i}{\pi}&\left(\ln\left(1-\frac{s_c}{a}\right)-\ln(s_c(a-s_c))-\ln\left(\frac{a+s_c}{a}\right)\right.\\
&\left.\;+\;\frac{2}{a}\sqrt{a^2-s_c^2}\ln(as_c)+\ln\left(-s_c(a+s_c)\right)\right)\\\\ 
\end{align*}
We can simplify the first logarithm, and cancel some factors.
\begin{align*}
\frac{s_c}{a}+\frac{i}{\pi}&\left(\ln\left(\frac{a-s_c}{a}\right)-\ln(s_c(a-s_c))-\ln\left(\frac{a+s_c}{a}\right)\right.\\
&\left.\;+\;\frac{2}{a}\sqrt{a^2-s_c^2}\ln(as_c)+\ln\left(-s_c(a+s_c)\right)\right)\\
\end{align*}
\begin{align*}
\frac{s_c}{a}+\frac{i}{\pi}&\left(\ln\left(\frac{1}{a}\right)-\ln(s_c)-\ln\left(\frac{a+s_c}{a}\right)\right.\\
&\left.\;+\;\frac{2}{a}\sqrt{a^2-s_c^2}\ln(as_c)+\ln\left(-s_c(a+s_c)\right)\right)\\\\ 
\end{align*}
We can split up some of these logarithms, and turn reciprocals into minus signs.
\begin{align*}
\frac{s_c}{a}+\frac{i}{\pi}&\left(-\ln\left(a\right)-\ln(s_c)+\frac{2}{a}\sqrt{a^2-s_c^2}\ln(as_c)\right.\\
&\left.-\ln\left(\frac{a+s_c}{a}\right)+\ln\left(s_c\right)+\ln\left(-1\right)+\ln\left(a+s_c\right)\right)\\\\ 
\end{align*} 
We can cancel the $\pm\ln\left(s_c\right)$ pair, and split up the logarithms further.
\begin{align*}
\frac{s_c}{a}+\frac{i}{\pi}&\left(-\ln\left(a\right)+\frac{2}{a}\sqrt{a^2-s_c^2}\ln(as_c)\right.\\
&\left.-\ln\left(a+s_c\right)+\ln\left(a\right)+\ln\left(-1\right)+\ln\left(a+s_c\right)\right)\\\\ 
\end{align*} 
We can also cancel the $\pm\ln\left(a\right)$ and $\pm\ln\left(a+s_c\right)$ pairs.
\begin{align*}
\frac{s_c}{a}+\frac{i}{\pi}&\left(\frac{2}{a}\sqrt{a^2-s_c^2}\ln(as_c)+\ln\left(-1\right)\right)\\\\ 
\end{align*}
The $\ln\left(-1\right)$ can be simplified to $\pi i.$ This can easily be derived from the equation $e^{i\pi}=-1.$
\begin{align*}
\frac{s_c}{a}+\frac{i}{\pi}&\left(\frac{2}{a}\sqrt{a^2-s_c^2}\ln(as_c)+\pi i\right)\\\\ 
\end{align*}
Distributing through the $i/\pi:$
\begin{align*}
\frac{s_c}{a}+&\frac{2i}{a\pi}\sqrt{a^2-s_c^2}\ln(as_c)-1\\\\ 
\end{align*} 
The real and imaginary parts are now separate. We know this because $a$ and $s_c$ are real, and $a>s_c,$ so $\sqrt{a^2-s_c^2}$ is real. Also, we showed earlier that $ab>0,$ so $\ln\left(as_c\right)$ must also be real. Therefore, the second term is completely imaginary, and the first term is completely real.
\begin{align*}
\frac{s_c-a}{a}+&\frac{2i}{a\pi}\sqrt{a^2-s_c^2}\ln(as_c)\\\\ 
\end{align*}

\subsubsection{Lower bound}\label{sec:appB:avg:lower}
Again, this is directly from Mathematica.
\begin{align*}\frac{2\sqrt{-s_c^2}\sqrt{1-\frac{s_c^2}{a^2}}\ln\left(a\sqrt{-s_c^2}\right)}{\sqrt{s_c^2}\pi}\\\\
\end{align*}
We know that $s_c>0$, so $\sqrt{s_c^2}\neq0,$ so we can cancel a factor of $\sqrt{s_c^2}.$
\begin{align*}\frac{2\sqrt{-1}\sqrt{1-\frac{s_c^2}{a^2}}\ln\left(a\sqrt{-s_c^2}\right)}{\pi}\\\\
\end{align*}
We can pull a factor of $a^{-1}$ out of the square root. Also, $\sqrt{-1}=i$
\begin{align*}
\frac{2i\sqrt{a^2 - s_c^2}\ln\left(a\sqrt{-s_c^2}\right)}{a\pi}\\\\ 
\end{align*} 
That logarithm can be pulled apart, and we can simplify $\sqrt{-s_c^2}$ to $s_c i.$
\begin{align*}
\frac{2i\sqrt{a^2 - s_c^2}\left(\ln\left(a\right)+\ln\left(s_c i\right)\right)}{a\pi}\\\\ 
\end{align*} 
The logarithm can be pulled apart further.
\begin{align*}
\frac{2i\sqrt{a^2 - s_c^2}\left(\ln\left(a\right)+\ln\left(s_c\right)+\ln\left(i\right)\right)}{a\pi}\\\\ 
\end{align*} 
We can recombine some logarithms, and the $\ln\left(i\right)$ can be simplified to $\pi i/2.$ This can easily be derived from the equation $e^{i\pi/2}=i.$
\begin{align*}
\frac{2i\sqrt{a^2 - s_c^2}\left(\ln\left(as_c\right)+\frac{\pi i}{2}\right)}{a\pi}\\\\ 
\end{align*} 
Distributing across the sum:
\begin{align*}
\frac{2i}{a\pi}\sqrt{a^2 - s_c^2}\ln\left(as_c\right)+\frac{2i}{a\pi}\sqrt{a^2 - s_c^2}\frac{\pi i}{2}\\\\ \end{align*} 
We can cancel some factors on the right.
\begin{align*}
\frac{2i}{a\pi}\sqrt{a^2 - s_c^2}\ln\left(as_c\right)-\frac{\sqrt{a^2 - s_c^2}}{a}\\\\ 
\end{align*} 
We have again separated the real and imaginary parts of this equation. The right is the same as last time, so we know that it is completely imaginary. The left must be real, because $\sqrt{a^2-s_c^2}$ is real, and $a$ is real.
\begin{align*}
-\frac{\sqrt{a^2 - s_c^2}}{a}+\frac{2i}{a\pi}\sqrt{a^2 - s_c^2}\ln\left(as_c\right)\\\\ 
\end{align*} 

\subsubsection{Difference of bounds}\label{sec:appB:avg:diff}
Subtraction should yield the definite integral, evaluated on $0\leq \cos i<b/a$
\begin{equation*}\frac{s_c-a}{a}+\frac{2i}{a\pi}\sqrt{a^2-s_c^2}\ln(as_c)-\left(-\frac{\sqrt{a^2 - s_c^2}}{a}+\frac{2i}{a\pi}\sqrt{a^2 - s_c^2}\ln\left(as_c\right)\right)\end{equation*}
Conveniently, the imaginary parts cancel cleanly, leaving only the real parts.
\begin{equation*}\frac{s_c-a}{a}-\left(-\frac{\sqrt{a^2 - s_c^2}}{a}\right)\end{equation*}
The difference of the real parts turns out to be rather simple:
\begin{equation*}\frac{s_c-a+\sqrt{a^2-s_c^2}}{a}\end{equation*}

\subsubsection{With $\cos i>\frac{s_c}{a}$ added in}\label{sec:appB:avg:final}
We must also remember to integrate from $s_c/a\leq\cos i\leq1.$ However, the time fraction usable here is just $1,$ which makes for easy integration.
\begin{equation*}\int_{s_c/a}^1 1\,d\left(\cos i\right) = 1-\frac{s_c}{a}\end{equation*}\newline
Adding that in with the other part of the integral:
\begin{equation*}\frac{s_c-a+\sqrt{a^2-s_c^2}}{a}+1-\frac{s_c}{a}=\frac{\sqrt{a^2-s_c^2}}{a}\end{equation*}\newline\newline
This gives us a simple, compact result.
\begin{equation}\label{eq:tfracavg}\begin{split}\lim_{\kappa\to\left(\frac{s_c}{a}\right)^{-}}&\int_0^{\kappa} \frac{2}{\pi}\cdot\arccos\left(\frac1a\cdot\sqrt{\frac{s_c^2-a^2 \cos^2i} {1-(\cos^2i)}} \right) d(\cos i)\\+&\int_{s_c/a}^1 1 \,d(\cos i)=\boxed{\frac{\sqrt{a^2-s_c^2}}{a}}\end{split}\end{equation}
\newline\newline

\section{Appendix C}\label{sec:appC}
\subsection{Inversion of $\frac{m}{x^2\sqrt{1-(n/x)^2}}$}
We need to invert the equation \begin{equation*}y=\frac{m}{x^2\sqrt{1-(n/x)^2}}\end{equation*}
We begin by squaring both sides:
\begin{align*}
    y^2&=\frac{m^2}{x^4\left(1-\frac{n^2}{x^2}\right)}\\
    &=\frac{m^2}{x^2\left(x^2-n^2\right)}
\end{align*}
We rearrange the equation:
\begin{align*}
    y^2&=\frac{m^2}{x^2\left(x^2-n^2\right)}\\
    x^2\left(x^2-n^2\right) &= \frac{m^2}{y^2}\\
    x^4-n^2x^2-\frac{m^2}{y^2}&=0
\end{align*}
We apply the quadratic formula and take the square root:
\begin{align}
    x^2 &= \frac{1}{2}\left(n^2\pm\sqrt{n^4+\frac{4m^2}{y^2}}\right)\nonumber \\\nonumber\\
    x &= \pm\sqrt{\frac{1}{2}\left(n^2\pm\sqrt{n^4+\frac{4m^2}{y^2}}\right)}\label{eq:quadinvert}
\end{align}

\bibliographystyle{aasjournal}
\bibliography{reference}
\appendix
\section{Index of variable names}\label{sec:varindex}
\begin{xltabular}{\linewidth}{ | l | X | l |}
    \hline\hline
    \textbf{\normalsize Variable Name} &\textbf{\normalsize Definition} & \textbf{\normalsize Units}\\ \hline\hline \endfirsthead
    \textbf{\normalsize Variable Name} &\textbf{\normalsize Definition} & \textbf{\normalsize Units}\\ \hline\hline \endhead
    $\eta_\oplus$ & The average number of earth-like planets hosted by a star. & Unitless\\\hline
    $D_{\lim}$ & The maximum distance away we will be looking. & Meters\\\hline
    $N_*$ & The number of stars that we will observe. & Unitless\\\hline
    $N_\oplus$ & The number of Exo-Earths that the survey aims to observe. & Unitless\\\hline 
    $\rho_*$ & The stellar density, assumed to be uniform. & cubic meter$^{-1}$\\\hline
    $\rho_\oplus$ & The density of Exo-Earths, also assumed to be uniform. & cubic meter$^{-1}$\\\hline
    $T$ & The total on-sky time. & Seconds\\\hline
    $R(\nu)$ & The rate at which a star emits photons of a given frequency. All stars are assumed to identical. & s$^{-1}$\\\hline
    $K$ & The ratio contrast ratio for the bandpass of interest. & Unitless\\\hline
    $R_e$ & The rate at which the telescope observes photons an the Exo-Earth. Equivalent to $\frac{RKd^2\varepsilon}{16D_k^2}$ & s$^{-1}$\\\hline
    $\varepsilon(f)$ & The laboratory efficiency, as a function of frequency. & Unitless\\\hline
    $\nu$ & The observational frequency. & Hertz\\\hline
    $D_k$ & The distance to the $k$th star. & Meters\\\hline
    $t_k$ & The amount of time spent on the $k$th star. $$T=\sum_{k=0}^{N_*} t_k$$ & Seconds\\\hline
    $d$ & The diameter of the telescope. & Meters\\\hline
    $SNR$ & The signal-to-noise ratio of the observation. Equivalent to $\sqrt{N_e}.$ Must be at least $SNR_0.$ & Unitless\\\hline
    $SNR_0$ & The minimum acceptable signal-to-noise ratio. & Unitless\\\hline
    $N_e$ & The number of photons detected by the telescope. & Unitless\\\hline
    $c$ & The total cost of the survey. & \$\\\hline
    $C$ & A scaling constant for the cost, such that $c=Cd^{2.5}$ & $\$/(\text{meters}^{2.5})$\\\hline
    $\text{iwa}$ & The inner working angle of the telescope. Approximated to be \begin{equation}\text{iwa}=\frac{n_i\lambda}{d}.\tag{\ref{eq:iwaApprox}}\end{equation} & Radians\\\hline
    $u_k$ & The amount of usable time spent on the $k$th star. & Seconds\\\hline
    $w_k$ & The amount of unusable time spent on the $k$th star & Seconds\\\hline
    $t_f$ & The fraction of time usable for each star. Depends on $a,$ $s_c,$ and $\cos i$ & Unitless\\\hline
    $a$ & The semi-major axis of an Exo-Earth. Determined by the location of the star's habitable zone. For solar analogues, this is close to Earth's semi-major axis. & Meters\\\hline
    $s_c$ & The projection of the inner-working-angle onto the sky. $D_k\cdot\text{iwa}$. & Meters\\\hline
    $\cos i$ & The cosine of a planet's orbital inclination. Assumed to be uniform random. & Unitless\\\hline
    $t_a$ & The average fraction of time usable, for a random $\cos i$. Defined as $$t_a=\int_{0}^{1} t_f\;\;d\cos i.$$ & Unitless\\\hline
    $\lambda$ & The observational wavelength. $\lambda=c/f.$ & Meters\\\hline
    $n$ & A group of variables, meant to simplify equations. Not $\eta_\oplus$-dependent. \begin{equation}n\equiv\frac{n_i\lambda}{a}\tag{\ref{eq:ndef}}\end{equation} & Meters$^{-1}$\\\hline
    $m$ & Another group of variables. Not $\eta_\oplus$-dependent. \begin{equation} m\equiv\frac{16SNR_0^2}{RK\varepsilon}\left(\frac{3}{4\pi\rho_*}\right)^{2/3}\tag{\ref{eq:mdef}}\end{equation} & Complicated\\\hline
    $x_p$ & The $x$-position of an exoplanet, from the viewer's perspective, as a function of time. $$x_p(t) = a\cos\left(t\right) $$ & Meters\\\hline
    $y_p$ & The $t$ position of an exoplanet, from the viewer's perspective, as a function of time. The $y$-axis is defined as the axis shortened by the exoplanet's inclination. $$y_p(t) = a\sin\left(t\right)\cos i $$ & Meters\\\hline
\end{xltabular}

\end{document}